\documentclass[fleqn,usenatbib]{mnras}
\usepackage{newtxtext,newtxmath}
\usepackage[T1]{fontenc}
\usepackage{ae,aecompl}
\usepackage{graphicx}	
\usepackage{amsmath}	
\newcommand{\rvec}{\mathbf{r}}
\newcommand{\vvec}{\mathbf{v}}
\newcommand{\evec}{\mathbf{e}}
\newcommand{\jvec}{\mathbf{j}}
\newcommand{\nvec}{\mathbf{n}}
\newcommand{\uvec}{\mathbf{u}}

\title[Scattering 1+2]{Strong Scatterings of Cold Jupiters and their Influence on Inner Low-mass Planet Systems: Theory and Simulations}

\author[Bonan Pu]{
 Bonan Pu$^1$ \thanks{E-mail: bp379@cornell.edu},
Dong Lai$^{1, 2}$ \thanks{E- mail: dong@astro.cornell.edu}
\\
1. Department of Astronomy, Center for Astrophysics and Planetary Science, Cornell University, Ithaca, NY 14853, USA \\
2. Tsung-Dao Lee Institute, Shanghai Jiao Tong University, Shanghai 200240, China
}

\pubyear{2020}

\begin{document}
\label{firstpage}
\maketitle

\begin{abstract}
Recent observations have indicated a strong connection between compact ($a \lesssim 0.5$ au) super-Earth and mini-Neptune systems and their outer ($a \gtrsim$ a few au) giant planet companions. We study the
dynamical evolution of such inner systems subject to the gravitational
effect of an unstable system of outer giant planets,
focussing on systems whose end configurations feature only a single
remaining outer giant. In contrast to similar studies which used on N-body simulations with specific (and limited) parameters or
scenarios, we implement a novel hybrid algorithm which combines N-body
simulations with secular dynamics with aims of obtaining analytical
understanding and scaling relations. We find that the dynamical evolution of the inner
planet system depends crucially on $N_{\mathrm{ej}}$, the number of mutual close
encounters between the outer planets prior to eventual
ejection/merger. When $N_{\mathrm{ej}}$ is small, the eventual evolution of the
inner planets can be well described by secular dynamics. For larger
values of $N_{\mathrm{ej}}$, the inner planets gain orbital inclination and eccentricity in a stochastic fashion analogous to
Brownian motion. We develop a theoretical model, and compute scaling laws for the final orbital parameters of the inner system. We show that our model can account for the observed eccentric super-Earths/mini-Neptunes with inclined cold Jupiter companions, such as HAT-P-11, Gliese 777 and $\pi$ Men.
\end{abstract}

\begin{keywords}
Dynamics - Extra-solar planets - Planetary Systems
\end{keywords}


\section{Introduction}
\label{sec:intro}

Exoplanets with masses and radii between that of the Earth and
Neptune, commonly referred to as ``super-Earths'' or
``mini-Neptunes'', have been discovered in large quantities in recent years.
Indeed, such planets appear to be ubiquitous in the Galaxy:
about $30\%$ of Sun-like stars host super-Earth planets, with each
system containing an average of 3 planets \citep{Zhu2018a}.
The observed super-Earth systems have compact orbits, with periods
typically less than 200 days. In recent years, an increasing number of
such systems have been found to host long-period giant planet
companions (i.e. ``Cold Jupiters'' or CJs).
\cite{Zhu2018b} analysed a sample of ground-based radial velocity
(RV) observations of super-Earth systems and an independent sample of
{\it Kepler} transiting Super-Earths with RV follow-up, and found that
cold Jupiters are three times more common around hosts of super-Earths
than around field stars: about $30\%$ the inner
super-Earth systems have cold Jupiter companions,
and the fraction increases to $60\%$ for metal-rich stars. \cite{Bryan2019} found a
similar result, and gave the estimated occurrence rate of $39\pm 7\%$ for
companions between 0.5-20$M_J$ and $1-20$~au.  There is evidence
that that stars with cold Jupiters or with high metallicities have smaller
multiplicity of inner Super-Earths, suggesting that cold Jupiters have influenced the inner planetary system. \cite{Masuda2020} found that these CJ companions are typically mildly misaligned with their inner systems with a mutual of $\Delta \theta \sim 12$ deg. These mild inner-outer misalignments could potentially explain the apparent excess of {\it Kepler} single-transit Super-Earth systems \citep{Lai2017}.

The question of how low-mass inner planet systems may be influenced by
the presence of one or more external giant planets has attracted recent attention 
\citep[e.g.][]{Carerra2016,Gratia2017,
  Lai2017,Huang2017, Mustill2017, Hansen2017, Becker2017, Read2017,
  JontofHutter2017, Pu2018, Denham2018}.
This paper is the third in a series where we systematically investigate the
effect of outer companions on the architecture of inner super-Earth
systems.  In \cite{Lai2017} and \cite{Pu2018}
we study the secular evolution of an inner multi-planet system
perturbed by an inclined and/or eccentric external companion. Combining 
analytical calculations and numerical simulations (based on secular 
and N-body codes), we quantify to what extent eccentricities and mutual inclinations
can be excited in the inner system for different masses and 
orbital parameters of super-Earths and cold Jupiter. 
When the perturber is sufficiently strong compared to the mutual gravitational
coupling between the inner planets, the inner system 
becomes dynamically hot and may be unstable.
Even for milder perturbers that do not disrupt integrity of the inner
system, the small/modest excitation of mutual inclinations can nevertheless disrupt the
co-transiting geometry of the inner planets and thereby reduce the
number of transiting planets \citep[e.g.][]{Brakensiek2016}. 
Other related works can be found in 
\cite{Boue2014, Hansen2017, Becker2017, Read2017, JontofHutter2017, Denham2018}
\cite[see also][for the effect of external companion on the stellar obliquity relative to the inner planets]{Boue2014b, Lai2018, Anderson2018}.

In this paper we study the dynamical evolution of inner planet systems
under the influence of a pair of external giant planets with initially
unstable orbits. A number of previous works (based on N-body
simulations) have already investigated this problem, illustrating that
the strong scatterings of unstable giant planets can affect the orbits
of the inner planets in different ways
\citep[e.g.][]{Matsumura2013, Carerra2016,Gratia2017,Huang2017, Mustill2017}.
For example, the outer
scatterings can send a giant planet inward, sweeping up all the inner
planets along its wake and totally destroying the inner system. Also,
the scattering events can excite the eccentricities and mutual
inclinations of the inner planets beyond the threshold of their
stability, causing the inner system to also undergo scattering events
of their own, resulting in a pared down inner system. In this paper we
attack this problem more systematically, going beyond previous works
in several ways. Our rationales are: (i) Previous works were
restricted to small number of numerical examples, often considering
specific orbital parameters. As such, it is difficult to obtain a
quantitative understanding or scaling relations (even approximate) in
order to know ``what systems lead to what outcomes''.  (ii) Previous
works often considered systems where the inner planets are not too
detached from the outer planets. This was adopted for numerical
reason: If the inner planets have too small a semi-major axis compared
to the outer planets, their dynamical times would be much shorter than
the outer planets, and it would be difficult to simulate the whole
system over a long time or simulate a large number of systems. As a
result, previous works tended to over-emphasize the more
``disruptive'' events. In reality, for sufficiently hierarchical
systems, the scattering events may only mildly excite the
eccentricities and mutual inclinations of the inner planets; in this
case, the super-Earths themselves are preserved, but their mutual
inclinations may be large enough to ``hide'' the inner planets from
simultaneously transiting their host stars -- such ``mild'' systems or
events may be most relevant to the currently observed super-Earths
with cold Jupiter companions.  (iii) Most importantly, there is a wide
range of ``ejection times'' associated with the evolution of the
unstable giant planets (e.g., for some systems, the lighter cold
Jupiter may be ejected very quickly, while for others the ejection may
take place over much longer time).  As we show in this paper, the
degree of influences on the inner system from the outer planets is
directly correlated with the ejection time of the unstable giant
planets. Thus, numerical studies that only consider restricted
examples would not capture the whole range of dynamical behaviors of
the ``inner planets + outer giants'' system.

Thus, the goal of this paper is to systematically examine how strong
scatterings of outer giant planets influence the inner super-Earth
system. We aim at obtaining an understanding of the whole range of
different outcomes and deriving relevant scaling relations for
different systems (with various planet masses and orbital parameters)
and different ejection times. Of particular interest are the ``mild''
systems where the inner planets survive the ``outer violence''.  We
elucidate the connections between the ``violent'' phase and the
ensuing ``secular'' phase studied in our previous papers \citep{Lai2017, Pu2018}. As mentioned above, because of the hierarchy of
dynamical timescales, it is difficult to study the systems where the
inner super-Earths and outer giants are well separated using
brute-force $N$-body simulations, especially when the ejection time of
giant planet is large -- and yet such systems are most relevant to the
observed super-Earths with cold Jupiter companions. To this end, we
developed a hybrid algorithm, combining $N$-body simulations of outer
giant planets undergoing strong scatterings with secular forcing on
the inner planets, to compute the evolution of the inner planets
throughout the ``violent'' phase.

A major part of this paper is devoted to the dynamics of strong
scatterings between two giant planets (Sec. \ref{sec:sec2}). Although there have been many previous studies on giant planet
scatterings \citep[e.g.][]{Rasio1996, Weidenschilling1996, Lin1997, Ford2000, Ford2008, Chatterjee2008, Juric2008, Matsumura2013, Petrovich2014, Frelikh2019, Anderson2020, li2020}, they all focused on the final outcomes of the unstable
giant planets (e.g., the eccentricity distribution of the remaining
planets), and did not investigate the timescale (``ejection time'') of violent phase.
As noted above, this ``ejection time'' directly influences the perturbations the inner 
planets receive from the ``outer violence''. In addition to obtaining 
the ``ejection time'' distribution, we also obtain a number of new analytic and scaling results for strong scatterings between two giant planets.

We then develop a theoretical model for the ``violent'' phase of the scattering
process, and model the inner planet's secular evolution as a linear
stochastic differential equation. We obtain analytic estimates for
both the expectation values and the distributions of the final orbital parameters of
the inner planets, and test these results against direct numerical integrations.
A major achievement of this work is the derivation for the marginalized
``violent-phase'' boost factor $\gamma$, which summarizes the entire
dynamics of the ``1+2'' scattering process in a single, dimensionless
parameter. We derive an analytical expression for the distribution of
$\gamma$, which agrees robustly with numerical simulations
over a wide range of initial system parameters.

This paper is structured as follows. In Sec. \ref{sec:sec2}, we
study the scattering process between two unstable giant planets using N-body simulations,
focusing in particular on the planet ejection timescale.
through N-body simulations. In Sec. \ref{sec:hybrid_algo}, we
outline our hybrid $N$-body and secular algorithm to study
the effect of giant planet scatterings on the inner super-Earth system.
In Sec. \ref{sec:1+2}, we present the results of these simulations, as well as theoretical scaling results for the final outcome of these systems. These results are extended to systems with more than one inner planets in Sec. \ref{sec:2+2}. We provide a summary of our results, and apply them to several ``Super-Earth + CJ'' systems of interest in Sec. \ref{sec:discussion}, as well as providing suggestions for further studies.

\section{Gravitational Scatterings of Two Giant Planets}
\label{sec:sec2}

The topic of gravitational scatterings between two or more giant
planets on unstable orbits is a classic one and has been the subject
of numerous previous studies \citep[e.g.][]{Rasio1996, Weidenschilling1996, Lin1997, Ford2000, Juric2008, Ford2008, Chatterjee2008, Ida2013,Matsumura2013, Petrovich2014, Frelikh2019, Anderson2020, li2020}. 
These studies focused on the final states of unstable systems, such as
the eccentricity distribution of the remaining planets.
We return to this topic to re-focus our attention on the
scattering/ejection timescale $t_{\rm ej}$, a quantity that plays a key
role in the interaction between the scattering CJs and the
inner super-Earth system, but hitherto ignored by previous studies
(but see Fig.~1 of \citealt{Anderson2020} and Fig.~7 of \citealt{li2020}). In particular, we seek to understand the distribution of $t_{\rm ej}$
and how the ejection outcome may scale with various system
parameters, such as the planet masses and spacing.
In this section, we present our numerical results (based
on $N$-body simulations) -- these empirical findings serve as the
basis for our theoretical model and analytical understanding discussed
in Section 3.

Consider a pair of planets with masses $m_1$ and $m_2$, radii $R_1$
and $R_2$ and semi-major axes $a_1$ and $a_2$ orbiting a star with
mass $M_{\star}$. We assume the planets are initially on circular
orbits and have a mutual inclination $0 < \theta_{12} \ll 1$ radians. The
planets are stable against close encounters for all time if the
condition
\begin{equation}
  |a_2-a_1| > 2\sqrt{3} r_H
\end{equation}
is satisfied \citep{Gladman1993},
where the mutual Hill radius $r_H$ is given by:
\begin{equation}
r_H \equiv \left(\frac{a_1 + a_2}{2}\right) \left(\frac{m_1 + m_2}{3M_*} \right)^{1/3}.
\end{equation}
If this condition is not satisfied, the resulting system is
gravitationally unstable and will inevitably undergo mutual close
encounters.  Generally, such
an unstable system will result in either the merger of two planets or
the ejection of one of the planets. The exact prevalence depends on
the initial system parameters, and planetary systems with smaller
semi-major axes and/or larger planetary radii are more likely to
result in collisions/mergers rather than planet ejections. For gas
giant planets with semi-major axes beyond a few au's, the most likely
outcome appears to be eventual ejection of the least massive planet
from the system. 
We focus on such ejection events in this section.

\subsection{Numerical Set-Up}
We perform N-body simulations of the orbital evolution of giant planets orbiting a solar mass star, using the IAS15 integrator included as part of the REBOUND N-body software \citep{Rein2012, Rein2015}. IAS15 is a 15th-order integrator based on Gauss-Radau quadruature with automatic time-stepping that is capable of achieving machine precision; it is well suited for problems involving close encounters and high-eccentricity orbits. 

We performed an array of N-body simulations involving the scattering of hypothetical unstable 2-planet systems. Each system had an inner planet with semi-major axis $a_1 = 5$ au, with the outer planet's
semi-major axis given by $a_2 = a_1 + k_0 R_H$, with $k_0 \in [1.5, 2.0, 2.5]$. The inner planet had mass $m_1 \in [10.0, 3.0, 1.0, 0.3]$ $M_{J}$ while the outer planet's mass is $m_2$, with the mass ratio $m_2/m_1$ chosen from [1, 2/3, 1/2, 1/3, 1/5, 1/10]. Note that in our simulations, the outer planet is less massive than the inner planet, although our analytic results apply to cases with the inner planet being more massive as well. The planets were treated as point particles (their radius were set to zero), and the possibility for collisions between planets were not considered. Both planets were started on initially circular orbits, and their initial orbital mutual inclination is set to be $\theta_{12,0} = 3^{\circ}$. The initial mean anomaly $f$, longitude of the ascending node $\Omega$ and longitude of pericenter $\varpi$ were each drawn from uniform distributions on $[0, 2\pi]$. We computed each system for up to $3\times 10^7$ orbits of the inner planet, terminating simulations once an ejection has occurred (i.e. the orbit of one of the planets becomes unbounded). For each combination of $k_0$, $m_1$ and $m_2/m_1$ we performed computations until 200 systems that resulted in ejected systems were obtained. The reason we perform such large numbers of simulations is to have sufficient data to test various statistical hypotheses that will arise later in the paper. The results of these simulations are summarized in the following sections.

\subsection{Final Outcomes of Scatterings: Orbital Parameters}
\label{sec:orbital_parameters}

After the scattering process has completed, we are interested in the final semi-major axis, eccentricity and inclination (relative to either the initial plane or the ejected planet) of remaining planet, which we denote as $a_{1,\mathrm{ej}}, ~e_{1,\mathrm{ej}}$ and $\theta_{1,\mathrm{ej}}$ respectively, with the subscripts ``$0$'' and ``ej'' denoting the quantity being at time zero and at the final time immediately after the ejection of the final planet. {\color{black} Although these results have been known and presented previously in various contexts (see references at the beginning of section 2), we explore a broader range of planets masses and mass ratios and test the
analytical scalings against simulations. }

\begin{enumerate}
\item{{\bf Final semi-major axis $a_{1,\mathrm{ej}}$}: {\color{black}The final semi-major axis is determined by the conservation of energy},
\begin{equation}
E_{\mathrm{tot}} = - \frac{GM_{\star}m_1}{2a_{1,0}} - \frac{GM_{\star}m_2}{2a_{2,0}} \simeq -\frac{GM_{\star}m_1}{2a_{1,\mathrm{ej}}},
\label{eq:e_conserv}
\end{equation}
which gives a final semi-major axis of 
\begin{equation}
a_{1,\mathrm{ej}} = a_{1,0} \left(1 + \frac{a_{1,0}m_2}{a_{2,0}m_1}\right)^{-1}
\label{eq:final_a}
\end{equation}
for the remaining, non-ejected planet.

{\color{black} In our simulations, we find that given the same set of initial planet masses and semi-major axes, the final distribution of the semi-major axis is determined by Eq. (\ref{eq:final_a}) to within $1 \%$}. This is a consequence of the diffusive nature of the ejection process, which proceed over many orbits through a series of energy exchanges, each exchange shifting the ejected planet's orbital energy by an amount $\delta E_{12} \ll E_{2,0}$. At ejection, the ejected planet deposits all its initial energy into planet 1, and the scatter in its final (positive) orbital energy is of order $\delta E_{12}$ and is negligible compared to the total energy lost $E_{2,0}$. }

\item{{\bf Final eccentricity $e_{1,\mathrm{ej}}$}: {\color{black}Our simulations show that the final eccentricity of the remaining planet depends strongly on the mass ratio $m_2/m_1$}, and weakly on the initial separation of the two planets. Figure \ref{fig:fig4} shows a plot of the distribution density of $e_1$ as a function of the mass ratio $m_2/m_1$ for a system with $m_1 = 1 M_J$. For $m_2 \ll m_1$ with initial separation of order $r_H$, a good empirical scaling for the typical value of $e_{1,\mathrm{ej}}$ is 
\begin{equation}
\langle e_{1,\mathrm{ej}} \rangle \approx 0.7 m_2/m_1.
\label{eq:final_e1}
\end{equation}
The spread in the value of $e_{1,\mathrm{ej}}$ increases with the mass ratio of the planet: for the case where $m_2/m_1 \ll 1$ (i.e. $m_2$ being a test particle), the standard deviation $\sigma(e_{1,\mathrm{ej}})$ is of order $\sim 0.25 \langle e_{1,\mathrm{ej}} \rangle$, while for the case of $m_2/m_1 \sim 0.5$ the standard deviation is $\sim 0.5 \langle e_{1,\mathrm{ej}} \rangle$. 

\begin{figure}
\includegraphics[width=0.95\linewidth]{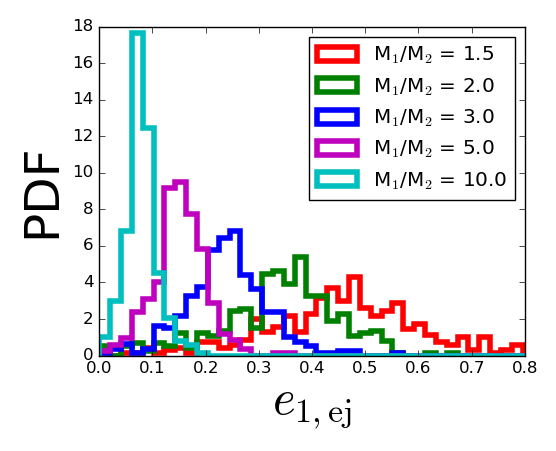}
\caption{A histogram of the final eccentricity of the remaining planet, for a system of two initial planets that have undergone an ejection event. The different colors correspond to various values of the mass ratio $m_1/m_2$. Each histogram represents 600 simulations, with $k_0 \in [1.5, 2.0, 2.5]$ (where $k_0 \equiv (a_2-a_1)/r_H)$ and $m_1 \in [3.0, 1.0]$ $M_{J}$. Runs with different $m_1$ were binned together as their distributions were indistinguishable statistically.}
\label{fig:fig4}
\end{figure}

The scaling of eccentricity can be understood as a consequence of the conservation of angular momentum:
\begin{equation}
m_1 \sqrt{GM_{\star}a_1(1-e_1^2)} + m_2 \sqrt{GM_{\star}a_2(1-e_2^2)} = \mathrm{const.}
\label{eq:L_conserv_de}
\end{equation}
We make the approximation that the apsis of the outer planet and the periapsis of the inner planet change much more slowly than their eccentricities and semi-major axes during close encounters, i.e.
\begin{align}
p_1 \equiv a_1(1+e_1) = a_{1,0} \simeq \mathrm{const.}
\label{eq:de1_de2_1} \\
q_2 \equiv a_2(1-e_2) = a_{2,0} \simeq \mathrm{const.}
\label{eq:de1_de2_2}
\end{align}
Combining Eqs. (\ref{eq:de1_de2_1}) - (\ref{eq:de1_de2_2}) with Eq. (\ref{eq:L_conserv_de}) and substituting a final value of $e_2 = 1$, we have
\begin{equation}
\sqrt{1 - e_{1,f}} \simeq 1 + (1-\sqrt{2})(m_2/m_1) \alpha_0^{-1/2},
\label{eq:e_1_gamma}
\end{equation}
where $\alpha_0$ is the initial value of the semi-major axis ratio $a_1/a_2$. In the limit that $(m_2/m_1) \ll 1$, Eq. (\ref{eq:e_1_gamma}) reduces to
\begin{equation}
e_{1,\mathrm{ej}} \approx 0.8 (m_2/m_1) .
\end{equation}
}

\item{{\bf Final inclination $\theta_{1,\mathrm{ej}}$}: We find $\theta_{1,\mathrm{ej}}$ to be determined most strongly by the mass ratio $m_2/m_1$, and somewhat independent of the other parameters. Fig. \ref{fig:fig5} shows our empirical results for the distribution of the inclination as a function of $m_2/m_1$. We find that $\theta_{1,\mathrm{ej}}$ is well-fit by a Rayleigh distribution with scale parameter $\sigma \sim 0.7\theta_{12,0}$. This can be understood as a consequence of angular momentum conservation. Since the ejected planet picks up a change in its angular momentum about the z-axis of order $\sin{\theta_{12,0}} L_{2,0}$, angular momentum conservation requires the remaining planet to gain angular momentum in equal and opposite direction. As a result, planet 1 will pick up an inclination relative to its original plane of order 
\begin{equation}
\theta_{1,\mathrm{ej}} \sim (L_{2,0}/L_{1,0}) \sin{\theta_{12,0}} \sim (m_2/m_1) \sin{\theta_{12,0}}.
\label{eq:final_t1}
\end{equation}

\begin{figure}
\includegraphics[width=0.95\linewidth]{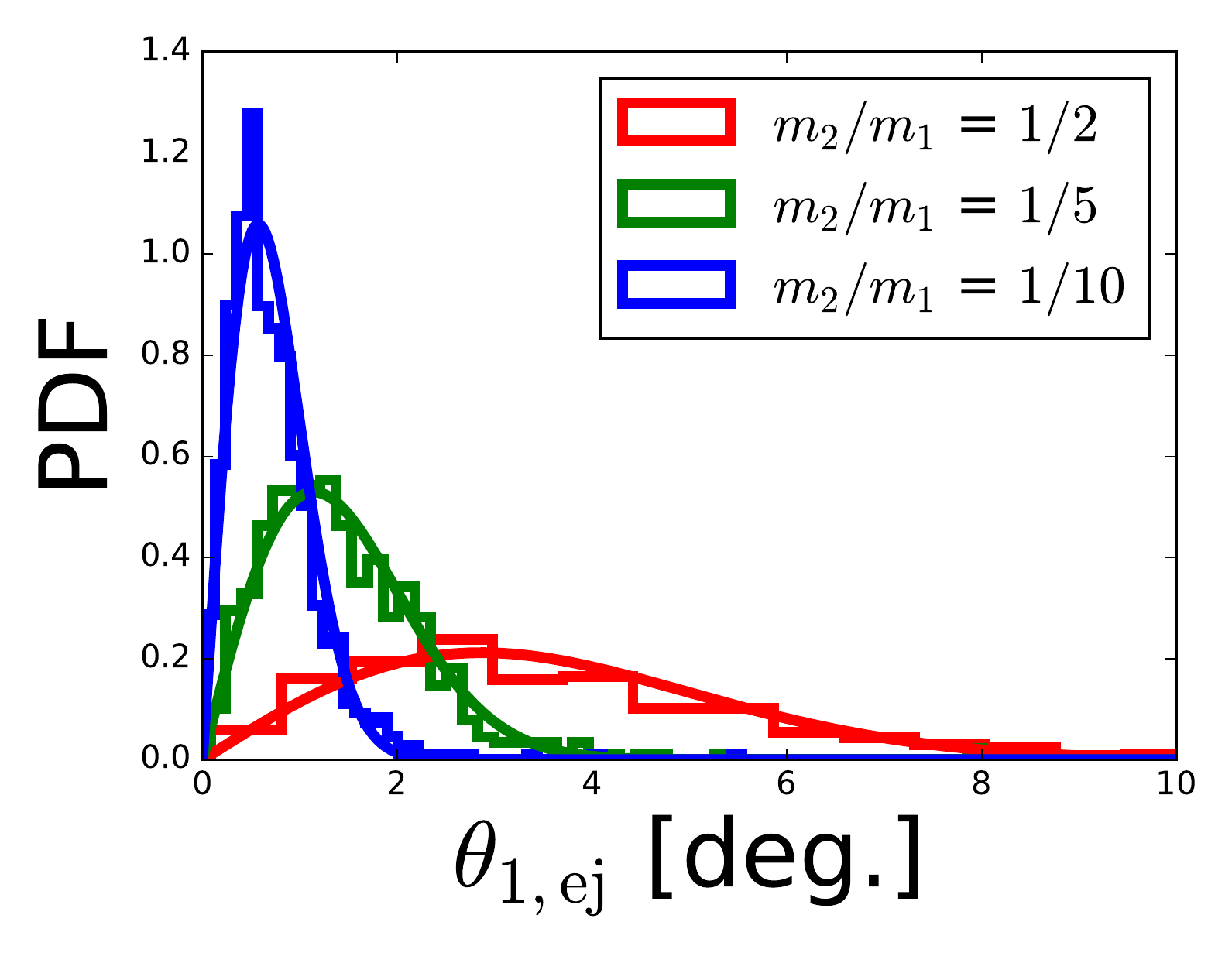}
\caption{A histogram of the final inclination of the remaining planet (relative to the initial plane), for a system of two initial planets that have undergone an ejection event. The initial mutual inclination of the two planets is $3^{\circ}$. The different colors correspond to various values of the mass ratio $m_2/m_1$. Each histogram represents 600 simulations, with $k_0 \in [1.5, 2.0, 2.5]$ and $m_1 = M_J$. Simulations with different $k_0$ were binned together as their distributions were approximately identical statistically.}
\label{fig:fig5}
\end{figure}
}
\end{enumerate}

\subsection{Timescale to Ejection}
\label{sec:Nej}
\begin{figure}
\includegraphics[width=0.95\linewidth]{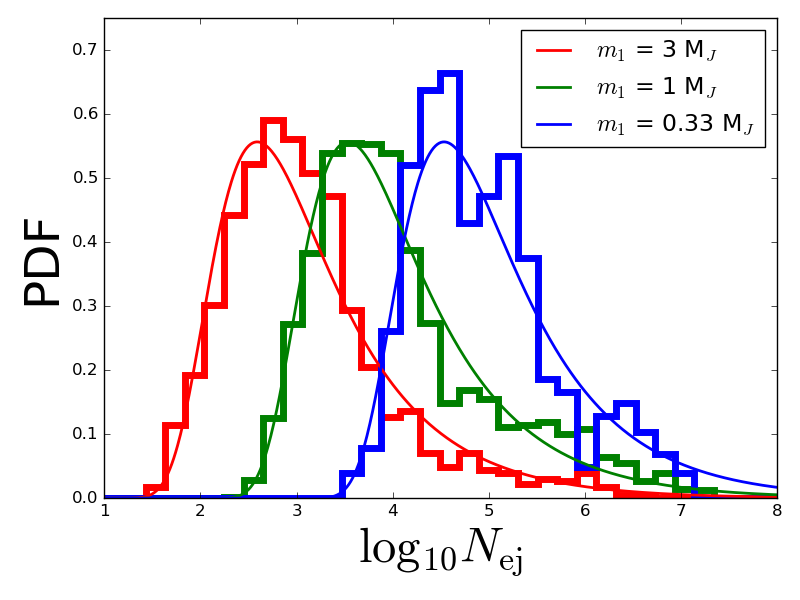}
\caption{Probability density distribution of $N_{\mathrm{ej}}$ from our two planet scattering simulations. The different colors represent different values of $m_1$, with red, green and blue corresponding to $m_1 = ~3, ~1, ~0.3 M_{\mathrm{J}}$ respectively. For each histogram, we fix $m_2/m_1 = 1/10$ and $k_0 = 2.0$. The histograms are empirical results from our N-body simulations, while the solid curves are obtained using the theoretical model in Eq. (\ref{eq:Nej_dist}), with $b$ empirically determined using Eq. (\ref{eq:b_MLE}). }
\label{Fig:Fig4}
\end{figure}

\begin{figure}
\includegraphics[width=0.95\linewidth]{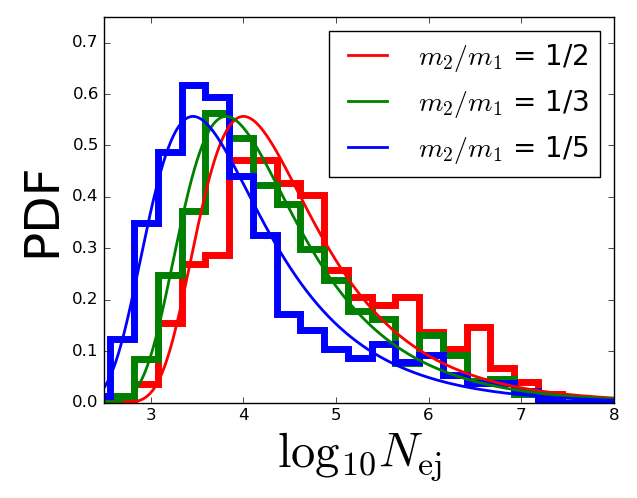}
\caption{Same as Fig. \ref{Fig:Fig4}, except we fix $m_1 = M_J$, while $m_2/m_1$ varies as indicated in the legend.}
\label{Fig:Fig5}
\end{figure}

\begin{figure}
\includegraphics[width=0.95\linewidth]{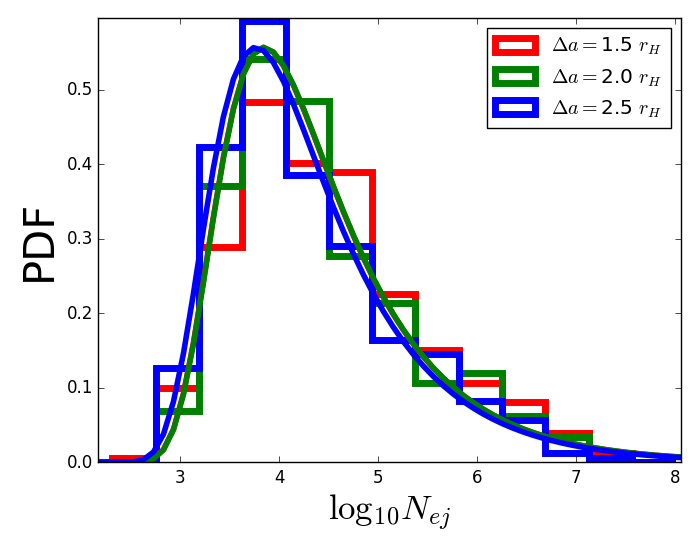}
\caption{Same as Fig. \ref{Fig:Fig4}, except we fix $m_1 = M_J$, while the initial separation parameter $k_0 = \Delta a / r_H$ varies as indicated in the legend.}
\label{Fig:Fig6}
\end{figure}

\begin{figure}
\includegraphics[width=0.95\linewidth]{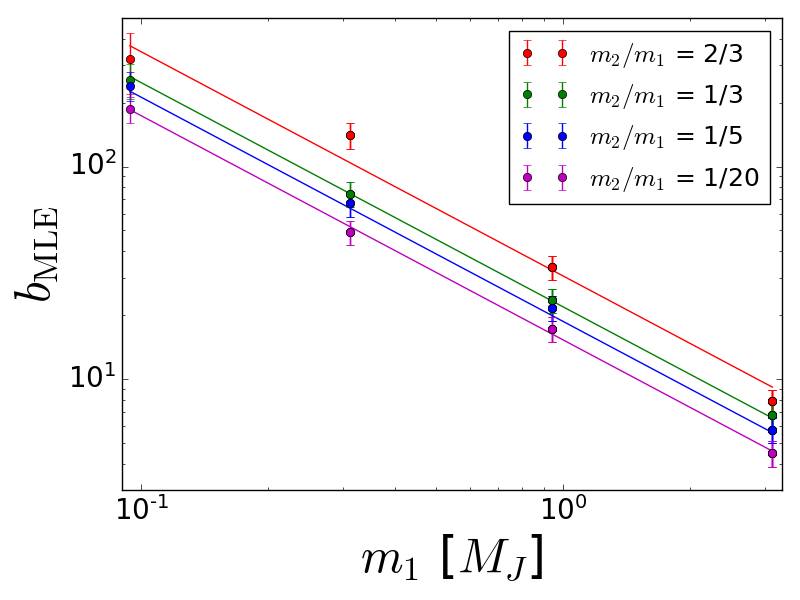}
\caption{The maximum likelihood estimate (MLE) estimate of $b$, as a function of $m_1$, for various combinations of the planet mass ratio $m_2/m_1$. The filled circles are the results of numerical N-body simulations, while the solid lines are given by Eq. (\ref{eq:b_fit}). The errorbars are computed using the asymptotic variance of the MLE (Eq. \ref{eq:b_var}).  }
\label{Fig:Fig7}
\end{figure}

An important quantity in the dynamical evolution of inner planet systems with scattering CJs is the timescale required to finally eject one of the planets. We present our empirical results on the scaling and dependence of the ejection timescale with system parameters. However, before proceeding, there are some caveats with regards to the correct metric to use for the ejection timescale.

Firstly, an unstable pair of planets on initially circular orbits will first pass through a meta-stable phase where the eccentricities of both planets ramp up gradually, without the planets under-going violent close encounters. This ramp-up phase is called the `instability timescale' $t_{\mathrm{inst}}$ in other contexts and its length depends on the parameters of the system. {\color{black} The scaling dependence of $t_{\mathrm{inst}}$ has been the subject of many studies \citep[e.g.][]{Chambers1996, Zhou2007, Smith2009, Pu2015, Obertas2017, Wu2019}, the results of which show that generally the instability timescale scales exponentially with the planet spacing, i.e. $\ln t_{\mathrm{inst}} \propto \Delta a$.  In this study we are interested in the timescale required for an initially unstable system to finally eject one of the planets, a process which only occurs after $t_{\mathrm{inst}}$ has already been reached \citep[see also][for a study on the timescale to the first planet-planet collision]{Rice2018}}. Therefore, it is convenient to separate the ramp-up phase from the ejection timescale by counting time only after the first close encounter. We do so by starting our count of the passage of time for planet ejections only after the planets 1 and 2 have orbits that are separated by a Hill radius or less, i.e. when $a_2(1-e_2) - a_1(1+e_1) \le r_H$ is satisfied.

We define $t_{\mathrm{ej}}$ and $N_{\mathrm{ej}}$ respectively as the time and the number of pericenter passages the ejected planet (planet 2) takes between the first Hill-sphere crossing event and the final ejection event. Note that we use the number of orbits of the ejected planet as opposed to the number of synodic periods, because at higher eccentricities the energy exchange mainly occurs at pericenter passages and not orbital conjunctions. $N_{\mathrm{ej}}$ and $t_{\mathrm{ej}}$ can be converted from each other using the transformations
\begin{align}
N(t) &\simeq \frac{1}{2\pi} \int_0^{t} \left(\frac{GM_*}{a^3(t)}\right)^{1/2} dt\\
t(N) &\simeq 2\pi \int_0^{N} \left(\frac{GM_*}{a^3(N)}\right)^{-1/2} dN.
\label{eq:t_to_N}
\end{align}
We focus on $N_{\mathrm{ej}}$ below, as it is the more physically relevant quantity in the scattering and ejection process. The results of of our numerical simulations are shown in Figs. \ref{Fig:Fig4} - \ref{Fig:Fig7}. We summarize the key results below:
\begin{enumerate}
\item{{\bf Dependence on $m_1$}: We find a strong dependence in our simulations of $N_{\mathrm{ej}}$ on the mass of the more massive planet $m_1$. The histograms in Fig. \ref{Fig:Fig4} show the different probability density distributions of $N_{\mathrm{ej}}$ for systems with various $m_1$ ranging from $3M_J$ to $0.3 M_J$. In our simulations, while systems with $m_1 = 3 M_J$ have $ N_{\mathrm{ej}} \sim 10^{3}$, the same system with a $m_1 = 0.3 M_J$ had a typical ejection timescale that is nearly a hundred times greater. We find that the scaling is very close to $N_{\mathrm{ej}} \propto m_1^{-2}$.}
\item {\bf Dependence on $m_2/m_1$}: For a given $m_1$, $N_{\mathrm{ej}}$ generally depends on $m_2/m_1$. When $m_2/m_1 \ll 1$, there is little dependence on $m_2$. On the other hand, as $m_2$ increases to be of similar order as $m_1$, the ejection timescale starts to increase significantly. Fig. \ref{Fig:Fig5} shows the density distribution of $N_{\mathrm{ej}}$ for a system with all other parameters fixed, except the ratio $m_2/m_1$, which is varied from $1/5 - 1/2$. We find that in comparison to the test-particle limit ($m_2/m_1 \ll 1)$, a mass ratio of $1/2$ results in an ejection timescale that is $\sim 10$ times larger. We find a scaling of $N_{\mathrm{ej}} \propto (1 + m_2/m_1)^{4.0}$, the functional form being somewhat arbitrary. 
\item{{\bf Variance of $N_{\mathrm{ej}}$}: in our simulations, we find significant variance in the distribution of $N_{\mathrm{ej}}$ for systems that have different initial orbital phases but otherwise identical orbital parameters. This can be seen clearly in Figs. \ref{Fig:Fig4} and \ref{Fig:Fig5}, where similar systems can have ejection timescales that range 4-5 orders of magnitude. We find that the standard deviation of $\log_{10}{N_{\mathrm{ej}}}$ is approximately $0.9$; this variance is empirically independent of the other system parameters such as planet masses.  }
\item{{\bf Dependence on $k_0$}: We found that the initial planet spacing $\Delta a = k_0 r_H$ plays little role in determining the final ejection timescale, as long as the initial ramp-up period of meta-stability is accounted for. Fig. \ref{Fig:Fig6} shows a comparison in the density distribution of $N_{\mathrm{ej}}$ for systems with otherwise identical parameters, except with $k_0$ varying from 1 to 2.5.}
\item{{\bf Relation between $N_{\mathrm{ej}}$ and ejection time $t_{\mathrm{ej}}$}: Since the semi-major axis of the planet increases as it is being ejected, the ejection time $t_{ej}$ is usually significantly larger than the naive estimate $t_{ej} \sim N_{ej} P_{2,0}$ where $P_{2,0}$ is the initial orbital period. The discrepancy grows larger when $m_1$ is smaller, due to the fact that the to-be-ejected planet can maintain larger semi-major axes before finally being ejected. We find a best-fit power-law with the form:
\begin{equation}
t_{\mathrm{ej}} \sim 8 P_{2,0} N_{\mathrm{ej}}^{0.7} \left(\frac{m_1}{M_{\star}}\right)^{0.46}.
\label{eq:t_ej_fit}
\end{equation}}
\end{enumerate}

\subsection{Theoretical Model for CJ Scattering}
We present a simple theoretical model for the process of CJ scattering to explain our empirical results of Section \ref{sec:Nej}. As we shall demonstrate in this section, by assuming that the planet orbital energy undergo a random walk during the scattering process, this model can explain both the distribution and the scaling of the ejection time of CJ scatterings.

Consider the limiting case of a pair of planets with $m_1 \gg m_2$. The two planet orbits are `unstable' such that their orbits come very close to each other and experience repeated crossings. At larger orbital distances it is common for the two planets to remain orbit-crossing for extended periods of time without physically colliding. Since $m_1 \gg m_2$, we assume the orbital parameters of $m_1$ stay constant during the scattering process.

At every pericenter passage (or apocenter passage if $a_2 < a_1$), planet 2 exchanges a certain amount of orbital energy with planet 1. The amount of energy exchanged, $\delta E_{12}$ depends on the orbital properties of the two planets. We hypothesize that $\delta E_{12}$ can be approximated as follows:
\begin{equation}
    \delta E_{12} \sim \left(\frac{G m_1 m_2}{a_1}\right)  F\left(a_2, ~f_{12}\right),
\end{equation}
where $F$ is a dimensionless function, and $f_{12}$ is the difference of the two planets' true longitudes at time of pericenter passage of planet 2. Note that in general, $F$ should depend on $e_2$ as well. However, given some $a_2$, the possible values of $e_2$ is narrowly constrained due to conservation laws (see Sec.  \ref{sec:inwards} below), so to a first order approximation, it is sufficient to know only $a_2$.

Due to symmetry, for a fixed value of $a_2$ the function $F$ is odd with respect to $f_{12}$, i.e. the energy exchange is equally likely to be positive and negative, and averaging over $f_{12}$ gives $\langle F(a_2, f_{12}) \rangle = 0$. As a result, even though at each close approach between planet 1 and planet 2 there is a finite amount of energy exchange, in the limit that $|\delta E_{12}| \ll E_{2}$, the long-term energy exchange is small, since $f_{12}$ is sampled almost periodically and uniformly. On the other hand, if $|\delta E_{12}| \sim E_{2}$, then each close encounter changes the period of planet 2 materially, such that the value of $f_{12}$ on the next approach is randomized. It is this randomization of the relative phase that causes energy exchange at iterative encounters to behave chaotically, resulting in a drift in orbital energy of planet 2 \cite[a similar phenomenon occurs when highly eccentric binaries experiences chaotic tides; see, e.g.][]{Vick2018}. 

In general, the amount of random diffusion in $E_2$ scales inversely proportional to the timescale in which the relative orbital phases $f_{12}$ at successive encounters can be randomized, so the energy exchange is most efficient at large values of $a_2$, and suppressed when $a_2$ is small. When eventually $E_2$ drifts to a positive value, the planet is ejected and the process terminates.

Now we study the question of for how long this process occurs, i.e. the mean value and distribution of $N_{\mathrm{ej}}$. To do this, we make use of a Brownian motion approximation in $E_2$ \cite[for a recent application of this idea in a different context, see][]{Mushkin2020}. 

Suppose we are able to find the RMS value of the function $F(a_2, f_{12})$ over the course of two-planet scattering, weighted by the likelihood of each $a_2$ occurring during the scattering process. We call this quantity $\bar{\delta}(m_1, m_2, a_1, a_{2,0})$, which depends on the initial separations, i.e.,
\begin{equation}
    \bar{\delta} \equiv \left(\frac{1}{2\pi}\int_{0}^{2\pi} \int_{0}^{\infty}  F^2(a_2, f_{12}) f(a_2) ~da_2 ~df_{12}\right)^{1/2},
\end{equation}
where $f(a_2)$ is the (unknown) probability density function of $a_2$ over the course of the scattering event. Then we may assume that the distribution of energy exchanges over the scattering process can be approximated as a Gaussian distribution with a mean of zero and width of $\bar{\delta}$. We do not attempt to compute $F(a_2, f_{12})$ or $f(a_2)$ explicitly; instead, we constrain them statistically from our N-body simulations by measuring the related parameter $b$, which is the ratio of the initial orbital energy and the RMS energy exchange and is given by
\begin{equation}
    b \equiv |E_{2,0}| \left(\frac{Gm_1m_2\bar{\delta}}{a_1}\right)^{-1}.
\end{equation}
In the limit of many successive passages, each giving a kick in energy that is small relative to the initial orbital energy $|E_{2,0}|$ (i.e. $N\gg 1$ and $b \gg 1$), the probability density distribution in $\Delta E_2/E_{2,0}$ after $N$ orbits is given by
\begin{equation}
    f(\Delta E_2/|E_{2,0}|) = \frac{1}{\sqrt{2\pi N}} \exp{\left(\frac{-(\Delta E_2/E_{2,0})^2}{2Nb^2}\right)}.
\label{eq:f_De}    
\end{equation}
$N_{\mathrm{ej}}$ is the lowest value of $N$ such that $\Delta E_2/|E_{2,0}| = 1$; it is known as the `stopping time' of the Weiner process and its probability density distribution is given by the Levy distribution \citep[see, e.g.][]{Borodin2002}:
\begin{equation}
f(N_{\mathrm{ej}}|b) = \frac{b}{\sqrt{2\pi N_{\mathrm{ej}}^3}} \exp{(-b^2/2N_{\mathrm{ej}})}.
\label{eq:Nej_dist}
\end{equation}
The distribution in Eq. (\ref{eq:Nej_dist}) is long-tailed since $f(N_{\mathrm{ej}})  \propto N_{\mathrm{ej}}^{-3/2}$ for $N_{\mathrm{ej}} \gg b^2$, and all of its moments including the arithmetic mean diverge. The geometric mean is $\langle N_{\mathrm{ej}}\rangle_{\mathrm{GM}} = \exp{(2\gamma_{\mathrm{EM}})} b^2 \approx 3.17b^2$ (where $\gamma_{\mathrm{EM}} \approx 0.57$ is the Euler-Mascheroni constant) and its mode is equal to $b^2/3$.  Another useful quantity is the harmonic mean, given by
\begin{equation}
    \langle N_{\mathrm{ej}}\rangle_{\mathrm{HM}} \equiv \langle 1/N_{\mathrm{ej}} \rangle^{-1} = b^2.
    \label{eq:nej_hm}
\end{equation}
The standard deviation of the quantity $\ln{N_{\mathrm{ej}}}$ is $\mathrm{Var}(\ln{N_{\mathrm{ej}}}) = \pi/\sqrt{2} \approx 2.2$, regardless of the value of $b$, and the 68\% and 95\% quantile ranges are $N_{\mathrm{ej}} \in [0.25b^2, 13b^2]$ and $[0.1b^2, 500b^2]$, respectively. In short, $N_{\mathrm{ej}}$ is distributed with a long tail at larger values and its distribution can easily span several orders magnitude.

The next step is to empirically determine the value of $b$ from the results of our numerical simulations, given the set of system parameters ($m_1, ~m_2, ~a_{2,0}$, etc.). To do so, we make use of the maximum likelihood estimate (MLE). The likelihood function for $K$ observations of $N_{\mathrm{ej,i}}, ~i \in [1, 2, ... K]$ is given by
\begin{equation}
   L(b) =  \prod_i^{K} \frac{b}{\sqrt{2\pi N_{\mathrm{ej,i}}^3}} \exp{(-b^2/2N_{\mathrm{ej,i}})}.
   \label{eq:Likelihood_Fn}
\end{equation}
Maximizing $\ln L$ with respect to $b$, we have
\begin{equation}
    {b}_{\mathrm{MLE}} = \underset{b}{\mathrm{argmax}} ~ L(b) = \sqrt{K} \left( \sum_i^{K} N_{\mathrm{ej},i}^{-1} \right)^{-1/2}.
    \label{eq:b_MLE}
\end{equation}
Its variance is given by the asymptotic variance of the MLE:
\begin{equation}
\mathrm{Var}({{b}_{\mathrm{MLE}}}) = \left(K \frac{\partial^2 L(b)}{\partial b^2}\right)_{b = {b}_{\mathrm{MLE}}}^{-1} = -2{b}_{\mathrm{MLE}}^2/K.
\label{eq:b_var}
\end{equation}

In Fig. \ref{Fig:Fig7} we show the empirical values of $b_{\mathrm{MLE}}$ estimated using Eq. (\ref{eq:b_MLE}) as functions of $m_1$ and $m_2/m_1$. We find that $b$ can be well-approximated by
\begin{equation}
    b \approx c_1 \left(\frac{m_1}{M_{\star}} \right)^{c_2} \left(1 + \frac{m_2}{m_1}\right)^{c_3}\left(\frac{a_{1,0}}{a_{2,0}} \right)^{c_4},
    \label{eq:b_fit}
\end{equation}
with $c_1 = 0.06 \pm 0.02$, $c_2 = -0.98 \pm 0.03$, $c_3 = 2.14 \pm 0.07$ and $c_4 = -1.4 \pm 0.5$; the above model has a value of $R^2 = 0.99$ when fitted against the empirical values of $b$ (as estimated by MLE). 

This empirical scaling is in fact consistent with the results of past studies, which showed that for comets with $a_2 \gg a_1$ and $(1 - e_2) \ll 1$, the RMS energy exchange per pericenter passage is of order $\delta E_{12} \sim Gm_1m_2/a_1$ \cite[see, e.g.][]{Wiegert1999, Fouchard2013}. This result would imply that $c_2 = c_4 = -1$, which is in agreement with our empirical results. 

Eqs. (\ref{eq:Nej_dist}) and (\ref{eq:b_fit}) provides an accurate description of the distribution for $N_{\mathrm{ej}}$ as long as $m_2/m_1 \lesssim 1/3$. However, this model breaks down in the comparable mass regime ($m_1 \sim m_2$), where $N_{\mathrm{ej}}$ is usually much larger than predicted by Eq. (\ref{eq:b_fit}). This is because for planets of comparable mass, as $a_2$ increases $a_1$ will decrease by a comparable value. As a result, the energy exchange becomes much less efficient as $a_2$ increases since the planet can only come close to one another when planet 1 and planet 2 are simultaneously at their apocenter and pericenter respectively. A theoretical model for this strong scattering process at comparable masses is an intriguing question in its own right, and necessary for further refinements on the results presented here, but beyond the scope of this paper.

\subsection{Scattering into inner system}
\label{sec:inwards}
Aside from the orbital parameters and ejection timescale, another quantity we are interested in is the minimum approach distance a planet might have with its host star. Since planet ejections occur gradually in a random walk-like manner, the ejected planet may first meander a significant amount inwards before being eventually ejected. If the to-be-ejected giant planet at some point comes too close to the inner system, it can undergo non-secular interactions with the inner system, causing our semi-secular approximation (see Section \ref{sec:hybrid_algo}) to break down. Therefore, it is important to quantify the extent to which the giant planet might first move inward.


First, due to conservation laws, there is a limit to how deeply inwards a planet can meander during the scattering process. If we assume the planet orbits remain (approximately) co-planar, then the 4 relevant variables are $a_1, ~a_2, ~e_1$ and $e_2$, which satisfy the constraints
\begin{itemize}
    \item Energy conservation: 
    \begin{equation}
        \sum_j m_j/a_{j,0}  = \sum_j m_j/a_j.
        \label{eq:constraint1}
    \end{equation}
    \item Angular momentum conservation: 
    \begin{equation}
        \sum_j m_j \sqrt{a_{j,0}(1 - e^2_{j,0})} = \sum_j m_j \sqrt{a_{j}(1 - e^2_{j})}.
        \label{eq:constraint2}
    \end{equation}
    \item Second law of thermodynamics: The system must not spontaneously `scatter' itself into a state that is indefinitely stable, even if this is permitted by the conservation laws. In general, the stability criterion for 2 planets with general masses, eccentricities and inclinations is complicated \citep[see, e.g.][]{Petrovich2015}. In the limit of co-planar orbits with $m_1 \gg m_2$, we find that requiring planets to follow the criterion below results in best agreement with the empirical results:
    \begin{equation}
        \frac{a_2(1-e_2)}{a_1(1+e_1)} \lesssim 1 + 2(m_1/3M_{\star})^{1/3}.
        \label{eq:constraint3}
    \end{equation}
    The above constraint asserts that the maximal planet separation should not exceed 2 Hill radii at all times.
\end{itemize}

The above three constraints reduce the degree of freedom to 1, which means that given any one variable, the other 3 variables are uniquely determined. One can then optimize for the lowest allowed values of $a_2$ and $a_2(1-e_2)$. This then produces a theoretical lower limit on $a_2$ during the scattering process. However, it is not a given that this minimum can always be reached, for two reasons: Firstly, since $\Delta E_2$ undergoes an approximate Brownian motion, it is likely to spend large fractions of time being positive, such that $a_2$ is never much below its initial value. Secondly, energy exchange becomes less efficient as $a_2$ decreases, since the timescale for the randomization of the relative orbital phase becomes larger. 

\begin{figure}
\includegraphics[width=1.01\linewidth]{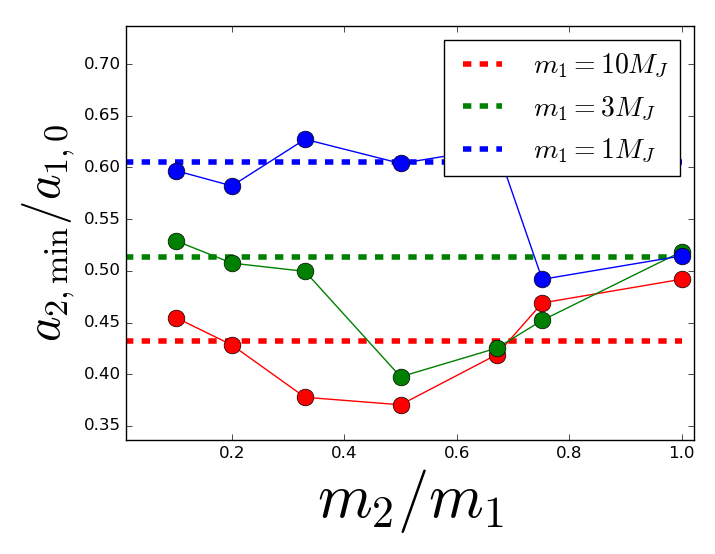}
\includegraphics[width=1.01\linewidth]{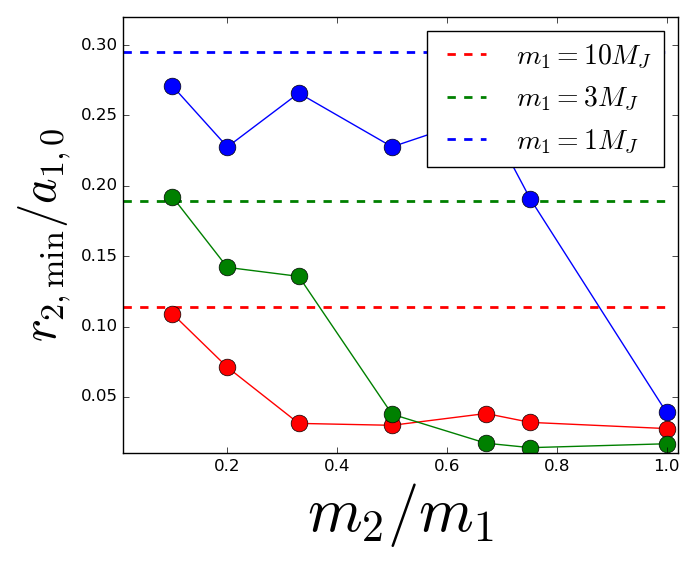}
\caption{Top: empirical values of $a_{2,\mathrm{min}}$ as function of $m_2/m_1$. Each data point represents the global minimum over all simulations. The blue, green and red circles correspond to $m_1 = 10, 3, ~1 M_J$ respectively. The dashed lines are derived from minimizing $a_{2}$ under the constraints given by Eqs. (\ref{eq:constraint1}) - (\ref{eq:constraint3}) in the limit of $m_2/m_1 \ll 1$. We suppress error bars in the empirical results because it is unclear how to estimate the minimum of a set of observations without prior assumptions about the distribution of our data. The bottom panel is similar to the upper panel, except we plot $r_{2, \mathrm{min}} = \min[{a_2(1-e_2)}]$ instead of $a_{2,\mathrm{min}}$. }
\label{Fig:inward_a}
\end{figure}

We show these limits for $a_2$ and $r_2 \equiv a_2(1-e_2)$, compared with empirical results from our simulations, in Fig. \ref{Fig:inward_a}. We see that generally, $a_{2,\mathrm{min}}/a_{1,0} \sim 1/2$, and decreases with increasing $m_1$. The theoretical constraints agreed well with empirical results when $m_2/m_1 \ll 1$, but breaks down when $m_2/m_1 \gtrsim 0.2$. We also find that $r_{2,\mathrm{min}}$ decreases strongly with increasing $m_2/m_1$, and can reach $r_{2,\mathrm{min}}/a_{1,0} \lesssim 0.05$ for $m_2 \sim m_1$. 

\begin{figure}
\includegraphics[width=1.01\linewidth]{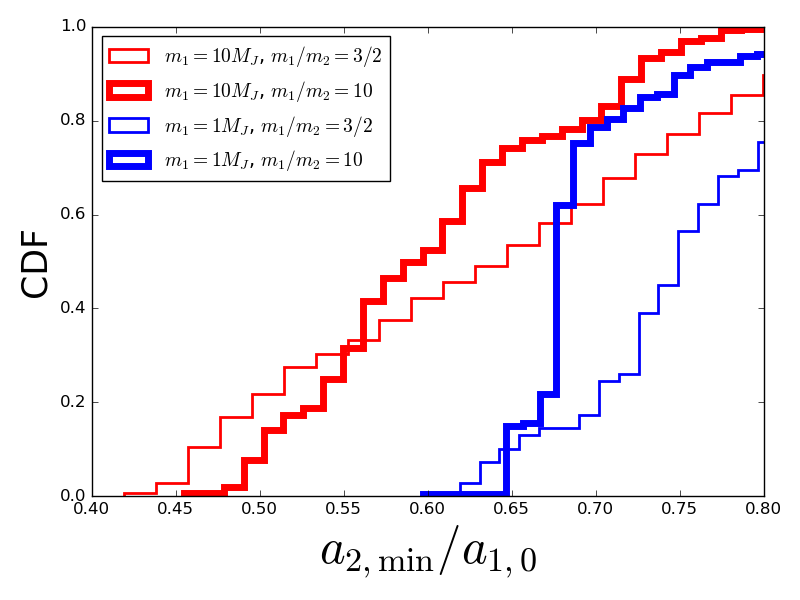}
\includegraphics[width=1.01\linewidth]{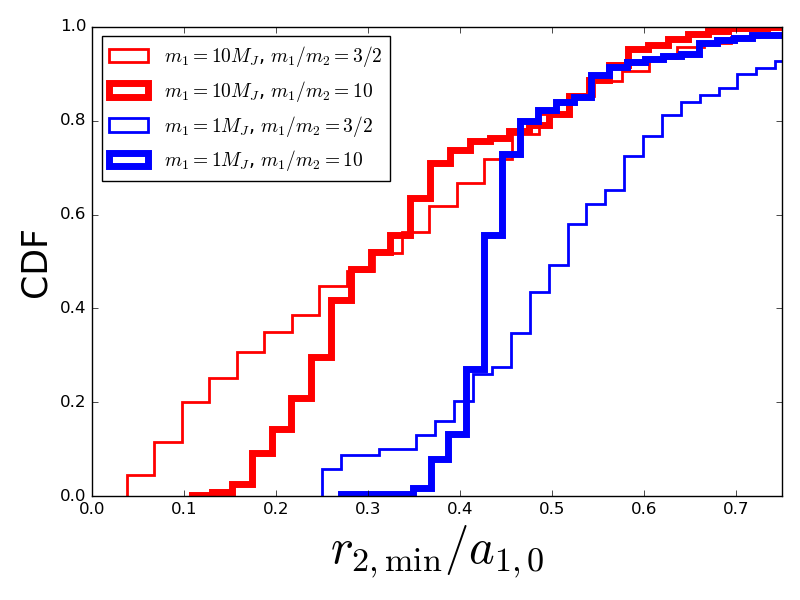}
\caption{Top: empirical cumulative distribution of of $a_{2,\mathrm{min}}$ as function of $m_1$ and $m_2/m_1$ from our sample of N-body simulations. The the red and blue histograms $m_1 = 10, ~1 M_J$ respectively. The thick lines correspond to $m_1/m_2 = 3/2$ while the thin lines correspond to $m_1/m_2 = 10$. Bottom panel: similar to the upper panel, except we plot $r_{2, \mathrm{min}} = \min[{a_2(1-e_2)}]$ instead of $a_{2,\mathrm{min}}$. }
\label{Fig:inward_a_hist}
\end{figure}

While it is possible for the less massive planet to be sent deep into the inner system during the scattering process when $m_2 \simeq m_1$, in practice this is an unlikely outcome. In Fig. \ref{Fig:inward_a_hist} we show the cumulative density distribution of realized $a_{2,\mathrm{min}}/a_{1,0}$ and $r_{2,\mathrm{min}}/a_{1,0}$ from our suite of N-body simulations. We find that $a_{2,\mathrm{min}}/a_{1,0}$ and $r_{2,\mathrm{min}}/a_{1,0}$ have a broad distribution: for $m_1 = 10M_J$ and $m_2 \simeq m_1$, $a_{2,\mathrm{min}}/a_{1,0}$ reaches below $1/10$ only $\sim 10\%$ of the time. When $m_1 \simeq m_2$, the empirical distribution for $a_{1,\mathrm{min}}/a_{1,0}$ and  $r_{1,\mathrm{min}}/a_{1,0}$ are very similar to $r_{2,\mathrm{min}}/a_{1,0}$ and $a_{2,\mathrm{min}}/a_{1,0}$ respectively, due to symmetry. Thus, even for initial parameters most likely to result in giant planets scattered deep into the inner system (i.e. $m_1 = 10M_J$ and $m_2 \simeq m_1$), the likelihood of one of the planets reaching a pericenter distance less than 1/10th of the initial semi-major axis is only $\sim 20\%$.

\section{Semi-Secular Algorithm for ``N+2'' Scatterings}
\label{sec:hybrid_algo}
We now consider how an inner low-mass planet system respond to an outer pair of giant planets undergoing strong scatterings. We label the inner planets as $j \in [a, b, c, ...]$, while the outer planets are labeled $p \in [1, 2]$. In Sec.  \ref{sec:1+2} we focus on inner systems with only one planet, and we extend our results to cases with 2 inner planets in Sec.  \ref{sec:2+2}, although our method can work for a general number of inner and outer planets. We imagine the inner system to be consistent with those discovered by Kepler, i.e. the planets have semi-major axes typically between $0.02 - 0.5$ au and are super-Earths in mass ($m_j \sim 3 - 20 M_{\oplus}$). We have a system of outer planets with semi-major axes beyond $\sim 2 - 3$ au that are gravitationally unstable ($k_0 \le 2\sqrt{3}$), and at least one of the planets have a fairly large mass ($\ge 100 M_{\oplus}$), although $m_2$ may be more comparable to super-Earths in size. We assume that the inner system is well-separated from the outer system ($a_j \ll a_1, ~a_2$), such that the inner planets do not participate directly in the outer scattering process.

As noted in Section \ref{sec:intro}, to address the question of how the inner planets are affected by the outer scattering, a direct approach based on N-body simulations is inadequate. The issue lies in the differing time-scales involved: The inner planets have short orbits on the timescale of days, which forces the time-step of the N-body simulation to not more than a few hours. On the other hand, the outer planets have periods of $\sim 10$ years and an ejection timescale of potentially hundreds of Myrs. To make matters even worse, the prospect of scattering events driven constantly by close encounters between planets preclude the use of fast and efficient symplectic integrators (e.g. the Wisdom-Holman mapping).

Here, we develop a hybrid method to evaluate the dynamical evolution of an inner system perturbed by a system of unstable outer CJs. In this method, we decouple the timescale of the inner planets and outer planets by computing their orbital evolutions separately. This is possible because we can safely neglect the back-reaction on the outer planets by the inners: since the inner planets are much less massive compared to their outer companions, the gravitational influence of the inner planets on the outer planets is negligible in comparison with the outer planets' own violent scatterings. Furthermore, since the inner planets are sufficiently far from the outer planets as to avoid direct scattering interactions, the gravitational influence by the outer planets is well described by secular dynamics \citep{Matsumura2013}.

Our algorithm is as follows. First, we evolve the gravitational interaction between the outer planets, in the absence of any inner planets. We then obtain a timeseries of the position-velocity vectors of each of the outer planets from beginning until final ejection. In the case of two giant planets, we have $\mathbf{r}_p(t)$ and $\mathbf{v}_p(t)$ for $p = 1, ~2$. These will be used as forcing terms to calculate the evolution of the inner planets, as follows.


Define $\jvec$ and $\evec$ as a planet's dimensionless angular momentum and eccentricity vectors:
\begin{equation}
{\bf j}=\sqrt{1-e^2}{\bf \hat n},\quad 
{\bf e}=e\,{\bf\hat u}
\end{equation}
where ${\nvec}$ and ${\uvec}$ are unit vectors, ${\nvec}$ is in the direction normal to the orbital plane and ${\uvec}$ is pointed along the pericenter. We compute the time evolution of these vectors for the outer planet $p$ using
\begin{align}
\jvec_p(t) &= \frac{1}{(GM_\star a_p)^{1/2} }\left[\rvec_p(t) \times \vvec_p(t) \right] \\
\evec_p(t) &= \frac{1}{GM_\star}\left[\vvec_p(t)\times (\rvec_p(t) \times \vvec_p(t)) \right].
\end{align}

According to Laplace-Lagrange theory \cite[e.g.][]{MurrayDermott}, the evolution equations for the eccentricity vector $\mathbf{e}_j$ and
unit angular momentum vector $\jvec_j$ on the planet $j$ due to the action of planet $k$, in the limit that $e_j$, $e_k$, $\theta_{jk}$ are small, are given by:
\begin{align}
\left(\frac{d \mathbf{e}_j}{dt}\right)_k &= -  \omega_{jk} (\mathbf{e}_j \times \jvec_k) +  \nu_{jk} (\mathbf{e}_k \times \mathbf{j}_j), 
\label{eq:LL_e}\\
\left(\frac{d \jvec_j}{dt}\right)_k &= \omega_{jk} (\jvec_j \times \jvec_k ).
\label{eq:LL_l}
\end{align}
The quantities $\omega_{jk}$ and $\nu_{jk}$ are the quadrupole and
octupole precession frequencies of the $j$-th planet due to the action
of the $k$-th planet, given by:
\begin{align}
\omega_{jk} = \frac{G m_j m_k a_{<}}{a_{>}^2 L_j } b^{(1)}_{3/2}(\alpha),
\label{eq:wjk} \\
\nu_{jk} = \frac{G m_j m_k a_{<} }{a_{>}^2 L_j} b^{(2)}_{3/2}(\alpha)
\label{eq:vjk}.
\end{align}
Here $a_< = {\rm min}(a_j,a_k)$, $a_> = {\rm max}(a_j,a_k)$, $\alpha = a_< / a_>$, 
$L_j \simeq m_j\sqrt{GM_*a_j}$ is the angular momentum of the $j$-th planet, and the $b^{(n)}_{3/2}(\alpha)$ 
are the Laplace coefficients defined by:
\begin{equation}
b^{(n)}_{3/2}(\alpha) = \frac{1}{2\pi} \int_{0}^{\pi} \frac{\cos{(nt)}}{(\alpha^2 + 1 - 2\alpha \cos{t})^{3/2}} dt.
\label{eq:laplace_coef}
\end{equation}


Laplace-Lagrange theory breaks down for more general values of $e_j$ and $\theta_{jk}$, and therefore, in this work we instead adopt a set of modified secular equations that interpolates between Laplace-Lagrange theory and secular multipole expansion. The equations are given in Eqs. (A2)-(A5) in \citep{Pu2018} and have better performance than Eqs. (\ref{eq:LL_e} - \ref{eq:LL_l}) when $e_j$ and $\theta_{jk}$ are large but $(a_a/a_1) \ll 1$. Thus we use these hybrid equations from \citep{Pu2018} in place of Eqs. (\ref{eq:LL_e}) - (\ref{eq:LL_l}) to compute the gravitational influence of the outer planets on the inner planets. Note that the adopted equations employ orbital averaging over both the inner planet and outer planet orbits. Even though the outer planet orbits vary on orbital timescales due to the strong mutual scatterings, the use of secular orbital averaging is appropriate since the interactions between the outer and inner planets are secular and accumulate over large number of orbits, the orbit-to-orbit variations can be ignored so long as the orbital period of outer planets is much shorter than the secular timescale.

In summary, we compute the evolution of the inner planets $j \in [a,b,c...]$, by the action of other inner planets $k \in [a,b,c...]$ as well as outer planets $p \in [1, 2]$ as follows:
\begin{align}
\frac{d{\jvec_{j}}}{dt} &= \sum_{k = a, b...} \left(\frac{d{\jvec_{j}}}{dt}\right)_k + \sum_{p = 1,2} \left(\frac{d{\jvec_{j}}}{dt}\right)_p,\\
\frac{d{\evec_{j}}}{dt} &= \sum_{k = a, b...} \left(\frac{d{\evec_{j}}}{dt}\right)_k + \sum_{p = 1,2} \left(\frac{d{\evec_{j}}}{dt}\right)_p.
\end{align}
The results of the calculations are discussed in Sec.  \ref{sec:1+2}.

\section{1+2 Scattering}
\label{sec:1+2}
We consider a single inner planet ("$a$") with two outer CJs. Planet $a$ has mass $3 M_{\oplus}$ and semi-major axis chosen from $a_a \in \{0.1, 0.15, 0.2, 0.25, 0.375, 0.5, 0.75, 1.0\}$, these are much smaller than the initial semi-major axes ($\ge 5$ au) of the outer planets so that planet $a$ typically does not participate directly in the scattering between planets 1 and 2. We assume all planets have initially circular and co-planar orbits, except that $\theta_{2,0} = 3$ degrees. We integrate this system using the semi-secular algorithm described in Sec.  \ref{sec:hybrid_algo}. A simulation is halted if any pair of planets undergo orbit crossings, or if planet $a$ attains an eccentricity greater than $0.99$. We discuss the results of these simulations below.
\subsection{Empirical Results}
\label{sec:1+2_empirical}
In our simulations we find a wide range of the final possible values of the inner planet eccentricity $e_a$, inclination $\theta_{a}$ measured relative to the original orbital plane of planet $a$ (note the orbits of planets $a$ and the remaining CJ are initially aligned), and mutual inclination $\theta_{a1}$ between the inner planet and the remaining CJ. As mentioned earlier, the evolution has two phases: the first phase is when the system has 3 planets total, with the outer two planets (planets 1 and 2) under-going scattering and the inner planet (planet $a$) interacting secularly with both planets. At some point, an outer planet is ejected, and the inner planet interacts with only the remaining CJ, whose orbital properties remain a constant in time. 

We define the eccentricity and inclination of the inner planet at the time of ejection as $e_{a,\mathrm{ej}}$ and $\theta_{a,\mathrm{ej}}$ respectively. After ejection, the inner planet still undergoes secular oscillations in eccentricity and inclination due to interactions with the remaining CJ. We thus define the time-averaged RMS eccentricity and inclination at infinity as
\begin{align}
    e_{a,\infty} &\equiv \left( \lim_{t \rightarrow \infty} \frac{1}{t - t_{\mathrm{ej}}} \int_{t_{\mathrm{ej}}}^{t} e^2_{a}(t) dt  \right)^{1/2}, \label{eq:e_infty}\\
    \theta_{a,\infty} &\equiv \left( \lim_{t \rightarrow \infty} \frac{1}{t - t_{\mathrm{ej}}} \int_{t_{\mathrm{ej}}}^{t} \theta_{a}^2(t) dt  \right)^{1/2}  \label{eq:th_infty}.
\end{align}
These quantities can be easily evaluated using secular theory \cite[see, e.g.][]{Pu2018}. For the mutual inclination, $\theta_{a1}$ remains constant once ejection has occured, thus $\theta_{a1, \infty} = \theta_{a1, \rm ej}$. Since the final value of $\theta_{1, \rm ej}$ is small (see Sec. \ref{sec:orbital_parameters}), in general $\theta_{a1, \infty} \approx \theta_{a, \infty}$. We focus on $e_{a,\infty}$ and $\theta_{a,\infty}$ as they are more representative of the long-term post-scattering dynamics of the inner planet. 

\begin{figure*}
\includegraphics[width=0.95\linewidth]{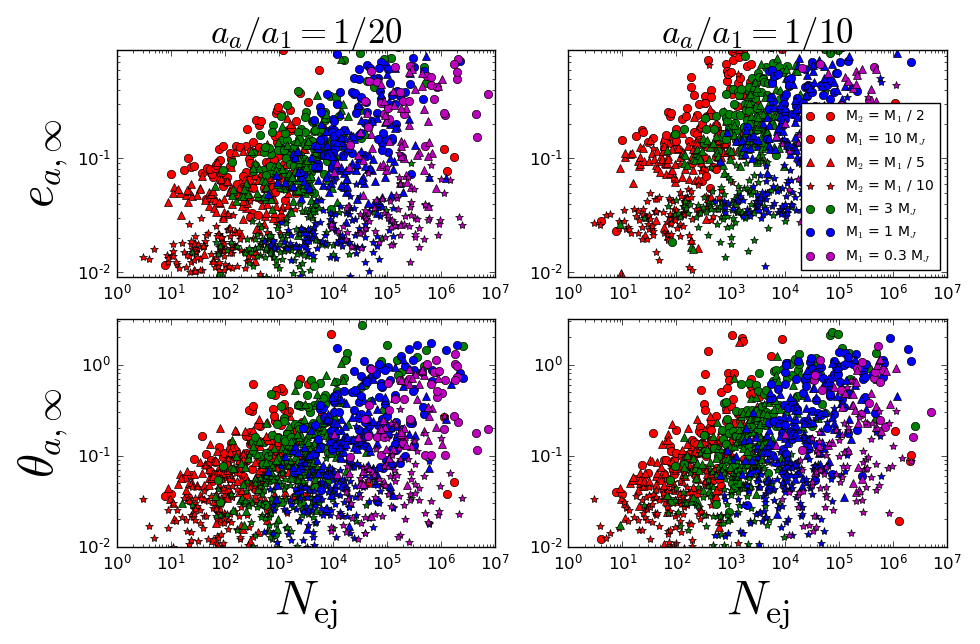}
\caption{The final values of $e_{a,\infty}$ (top panels) and $\theta_{a,\infty}$ (in radians, bottom panels) as defined by Eqs. (\ref{eq:e_infty}) - (\ref{eq:th_infty}), as a function of $N_{\mathrm{ej}}$, for a 1-planet inner system subject to the gravitational influence of two scattering giant planets. The masses of the outer planets are varied with $m_1 = 10, ~3, ~1$ or $0.3 M_J$ (the red, green, blue and magenta points respectively), while the mass ratio $m_2/m_1 = 1/2, ~1/5, ~ 1/10$ for the filled circles, triangles and stars respectively. The initial semi-major axes of the outer planets are $a_1 = 6.0$ au and $a_2 = a_1 + k_0 r_H$ with $r_H$ being the mutual Hill radius and $k_0$ chosen randomly from [1.5, 2.0, 2.5]; the value of $k_0$ matters little for the final results. The left panels show systems where the initial $a_a/a_1 = 1/20$, while the right panels have $a_a/a_1 = 1/10$.}
\label{Fig:Fig11}
\end{figure*}

Fig. \ref{Fig:Fig11} shows the values of $e_{a,\infty}$ and $\theta_{a,\infty}$ for a subset of our simulations. According to Fig. \ref{Fig:Fig11}, $e_{a,\infty}$ and $\theta_{a,\infty}$ tends to increase roughly as $\sqrt{N_\mathrm{ej}}$. We provide a theoretical model for this behavior in Sec.  \ref{sec:simple_model_1+2}. Secondly, we find a strong dependence of the final values of $e_{a,\infty}$ and $\theta_{a,\infty}$ on the planet mass ratio $m_2/m_1$, with outer planet pairs having comparable masses leading to much higher values of $e_{a,\infty}$ and $\theta_{a,\infty}$ compared with cases where $m_1 \gg m_2$. The main reason is that these final values increase as the mass ratio $m_2/m_1$ increases, and more eccentric/inclined perturbers tend to drive stronger perturbations on the inner planet. 

\begin{figure*}
\includegraphics[width=0.95\linewidth]{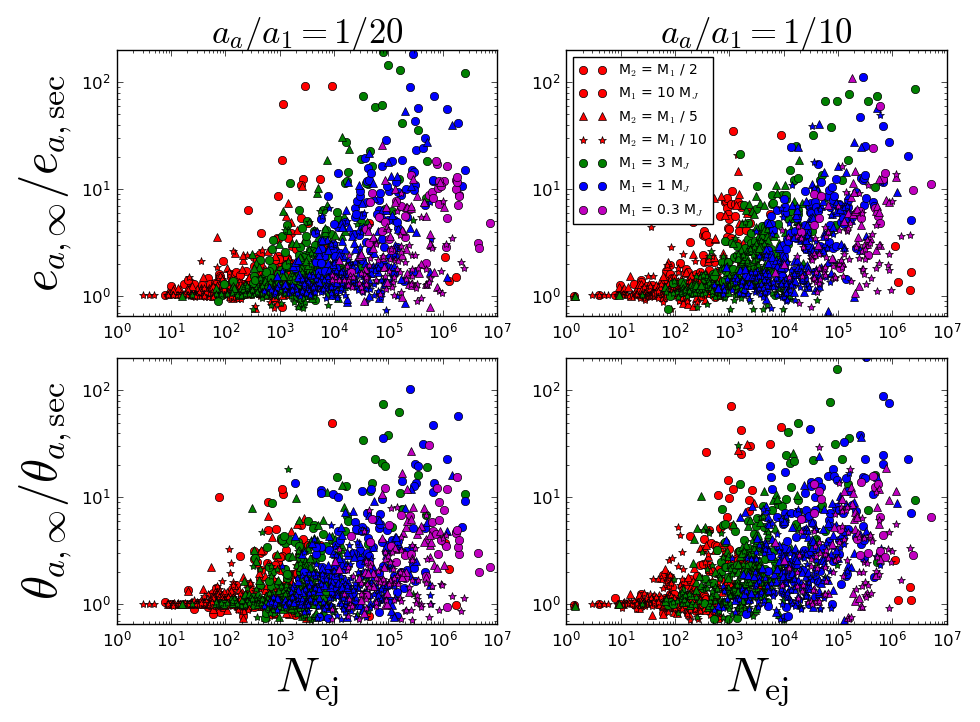}
\caption{Same as the Fig. \ref{Fig:Fig11}, except the eccentricities and inclinations are normalized by the ``secular'' expectation $e_{a,\mathrm{sec}}$ and $\theta_{a,\mathrm{sec}}$ given by Eqs. (\ref{eq:e_sec}) - (\ref{eq:t_sec}). }
\label{Fig:Fig12}
\end{figure*}
How to understand the diversity of final results in this parameter space? The picture becomes clearer if we normalize the results by the ``scattering-free'' theoretical expectations. We introduce these ``scattering-free'' quantities as the ``secular'' eccentricity and inclination $e_{a,\mathrm{sec}}$ and $\theta_{a,\mathrm{sec}}$ that are the (RMS) eccentricities and inclinations that would be expected on planet $a$, if the the dynamical history of the two-planet scattering were to be ignored, and the inner planets started their orbital evolution with $m_1$ at its final orbital state and $m_2$ removed. In other words, $e_{a,\mathrm{sec}}$ and $\theta_{a,\mathrm{sec}}$ are RMS eccentricity and inclination that planet ``a'' would finally obtain, if it started on an initially circular, non-inclined orbit under the influence of the perturber planet ``1'' with initial eccentricity and inclination $e_1 = e_{1,\mathrm{ej}}$, $\theta_{a} = \theta_{a,\mathrm{ej}}$. For $L_a \ll L_1$, we have \citep[e.g.][]{Pu2018}:
\begin{align}
e_{a,\mathrm{sec}} &= \frac{5\sqrt{2} a_a e_{1,\mathrm{ej}}}{4a_1(1-e_{1,\mathrm{ej}}^2)} 
\label{eq:e_sec}\\
\theta_{a, \mathrm{sec}}  &= \sqrt{2} \theta_{1,\mathrm{ej}}
\label{eq:t_sec}\\
\theta_{a1, \rm sec} &= \theta_{1, \rm ej}
\label{eq:ta1_sec}
\end{align}
(note that $\theta_{1,\mathrm{ej}}$ is the inclination of planet $1$ measured relative to its initial orbital plane). Fig. \ref{Fig:Fig12} shows our numerical results of Fig. \ref{Fig:Fig11} for the final RMS values of $e_{a,\infty}$ and $\theta_{a,\infty}$, normalized by the secular expectations $e_{a,\mathrm{sec}}$ and $\theta_{a, \mathrm{sec}}$. We find that the scaling for the final values of $e_{a,\infty}$ and $\theta_{a,\infty}$ can be divided into two regimes. In the case where $N_{\mathrm{ej}}$ is small, $e_{a,\infty}$ and $\theta_{a,\infty}$ reduce to their ``secular'' expectations. In the case that $N_{\mathrm{ej}}$ is large, the ratio $e_{a,\infty} / e_{a,\mathrm{sec}}$ and $\theta_{a,\infty}/ \theta_{a, \mathrm{sec}}$ can be either larger or smaller than 1, and is bounded from below by $\sqrt{2}/2$; the average values scale proportionally to $\sqrt{N_{\mathrm{ej}}}$, albeit with a large spread. The transition between the two regimes occur approximately at $N_{\mathrm{ej}} \sim N_{\mathrm{sec}}$, with $N_{\mathrm{sec}}$ given by
\begin{equation}
N_{\mathrm{sec}} \equiv \left(\frac{\omega_{a1,0}P_{1,0}}{ 2\pi} \right)^{-1} = \frac{1}{2\pi} \left(\frac{m_1}{M_{\star}} \right)^{-1} \left(\frac{a_a}{a_1} \right)^{-3/2} ,
\label{eq:Nsec}
\end{equation}
where $\omega_{a1,0}$ is the (initial) secular quadrupolar precession frequency of planet $a$ driven by planet 1 (see Eq. \ref{eq:wjk}) and $P_{1,0}$ is the initial orbital period of planet 1. This boundary is consistent with the inner planet $a$ being driven by stochastic secular forcing from planets 1 and 2 during the ejection process: When $N_{\mathrm{ej}} \ll N_{\mathrm{sec}}$, the ejection occurs much more quickly than the timescale of secular interactions, and the dynamical history of the ejection can be ignored. On the other hand, when $N_{\mathrm{ej}} \gg N_{\mathrm{sec}}$, the stochastic `forcing' on planet $a$ driven by the scattering perturbers will cause $e_a$ and $\theta_{a}$ to undergo a random walk of its own, with the value of $e_{a,\infty}$ and $\theta_{a,\infty}$ scaling proportionally to $\sqrt{N_{\mathrm{ej}}}$. 

\begin{figure}
\includegraphics[width=0.95\linewidth]{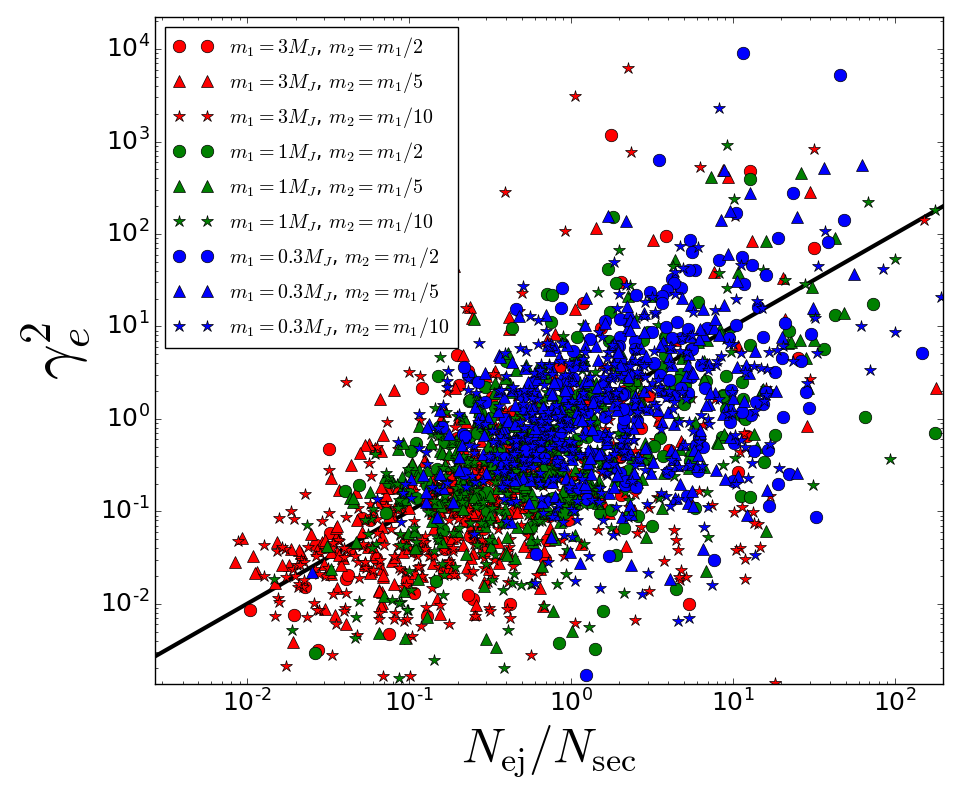}
\caption{The value of $\gamma_e^2$ (Eq. \ref{eq:gamma_e}) plotted as a function of $N_{\mathrm{ej}} / N_{\mathrm{sec}}$ (see Eq. \ref{eq:Nsec}) for our simulations. Here $a_a = 0.3$ au (corresponding to $a_a/a_1 = 1/20$). Red, green and blue points correspond to $m_1 = 3, ~1, ~0.3 M_J$ respectively. The filled circles, triangles and stars correspond to $m_2/m_1 = 1/2, ~1/5, ~1/10$ respectively. The black solid line is given by $\gamma_e^2 = N_{\mathrm{ej}}/N_{\mathrm{sec}}$. }
\label{Fig:Fig8}
\end{figure}

\begin{figure}
\includegraphics[width=0.95\linewidth]{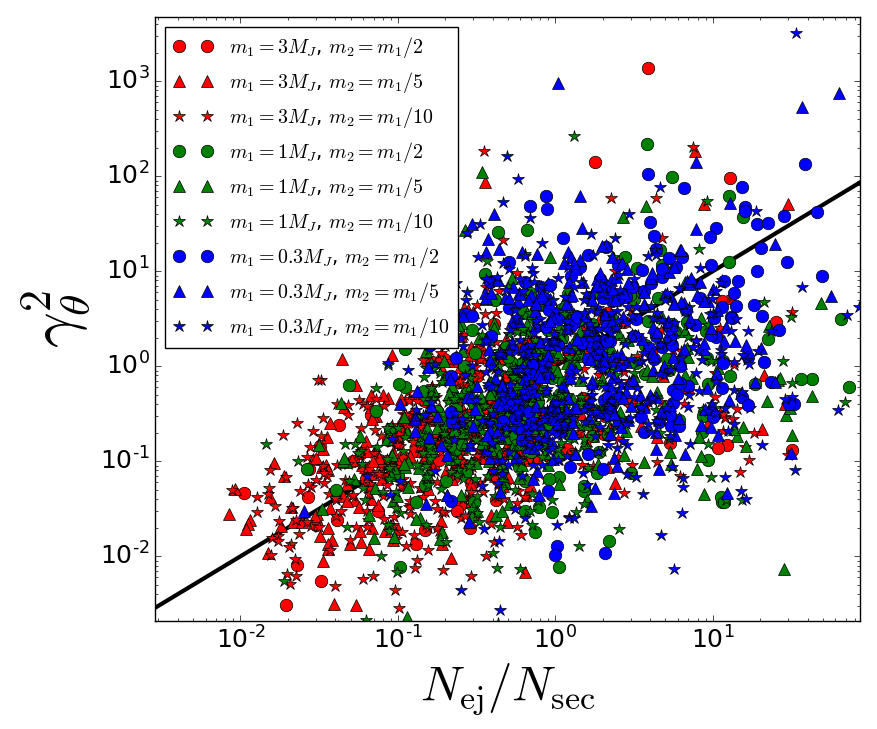}
\caption{Same as Fig. \ref{Fig:Fig8}, except we show $\gamma^2_{\theta}$ as defined by Eq. (\ref{eq:gamma_t}).}
\label{Fig:Fig9}
\end{figure}

The final results can be summarized most succinctly if we consider the deviation of the final values of $e_a$ and $\theta_{a}$ from their secular predictions and define the ``boost factors'':
\begin{align}
\gamma_e^2 &\equiv \frac{|e^2_{a,\infty}  -  e^2_{a,\mathrm{sec}}|}{  e^2_{a ,\mathrm{sec}}}
\label{eq:gamma_e} \\
\gamma_\theta^2 &\equiv \frac{| \theta_{a, \infty}^2  - \theta_{a,\mathrm{sec}}^2 |}{\theta_{a,\mathrm{sec}}^2 }
\label{eq:gamma_t}.
\end{align}
Figs. \ref{Fig:Fig8} and \ref{Fig:Fig9} show the comparison of our numerical results for the values of $\gamma_e$ and $\gamma_{\theta}$ for a subset of our numerical integrations. We find that across a wide range of parameters for $a_a$, $a_1$, $m_a$, $m_1$ and $m_2$, the quantities $\gamma_e, ~\gamma_{\theta}$ have a universal scaling given by (shown as the solid black line in Figs. \ref{Fig:Fig8} and \ref{Fig:Fig9}):
\begin{equation}
\gamma_e \sim \gamma_{\theta} \sim \sqrt{N_{\mathrm{ej}}/N_{\mathrm{sec}}}.
\label{eq:dynsec_scaling}
\end{equation}
The boost factor for the mutual inclination, defined as
\begin{equation}
\gamma^2_{\theta, a1} \equiv \frac{| \theta_{a1, \infty}^2  - \theta_{a1,\mathrm{sec}}^2 |}{\theta_{a1,\mathrm{sec}}^2} \label{eq:gamma_a1}
\end{equation}
also shows the same scaling, but with different normalization. We find that $\gamma_{\theta,a1} \sim 1.4 \gamma_{\theta}$; we provide a theoretical explanation for this in Sec. \ref{sec:inclination}.

\begin{figure}
\includegraphics[width=0.95\linewidth]{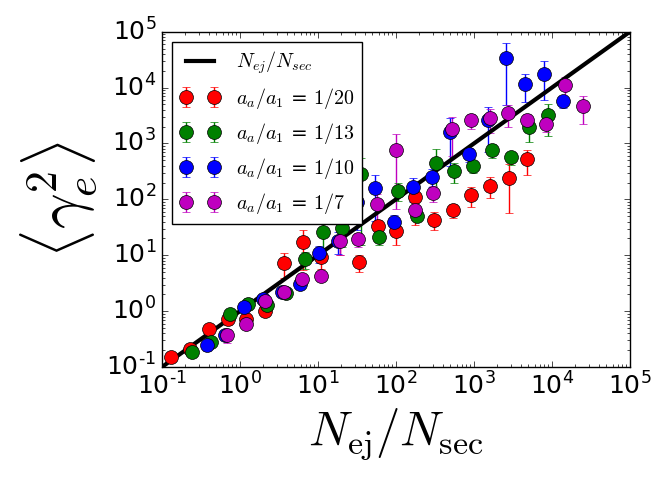}
\caption{The average value of $\gamma_e^2$, binned by $\log{(N_{\mathrm{ej}}/N_{\mathrm{sec}}}$ with 4 bins per logarithmic decade, as a function of ${N_{\mathrm{ej}}/N_{\mathrm{sec}}}$. For each of the points, $m_1 = M_J$ and $m_2/m_1 = 1/5$. The red, green, blue and magenta filled circles correspond to $a_a/a_1 = 1/20, 1/13, 1/10$ and $1/7$ respectively. The errorbars are given by the standard error, and the solid black line is given by $\langle \gamma_e^2 \rangle = N_{\rm ej}/N_{\rm sec}$.}
\label{Fig:gamma_e_by_aa}
\end{figure}

\begin{figure}
\includegraphics[width=0.95\linewidth]{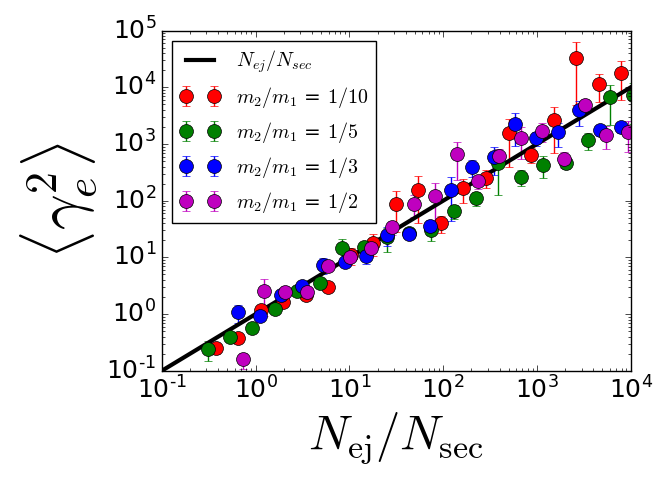}
\caption{Same as Fig. \ref{Fig:gamma_e_by_aa}, except that we fix $a_a/a_1 = 1/10$, and $m_2/m_1$ varies as indicated by the plot legend. }
\label{Fig:gamma_e_by_mr}
\end{figure}

\begin{figure}
\includegraphics[width=0.95\linewidth]{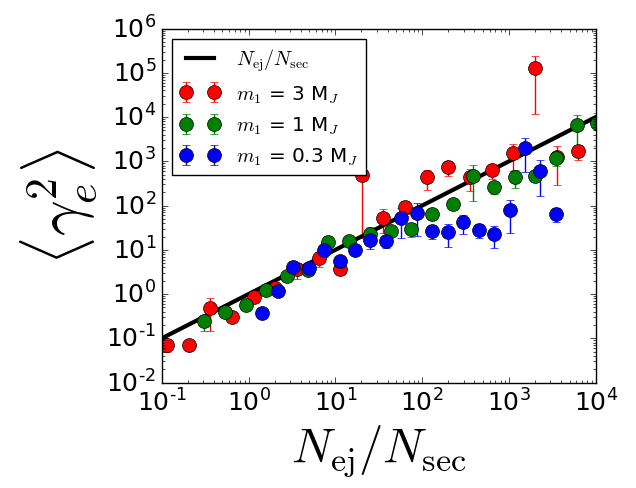}
\caption{Same as Fig. \ref{Fig:gamma_e_by_aa}, except that we fix $a_a/a_1 = 1/10$, and $m_1$ varies as indicated by the plot legend. }
\label{Fig:gamma_e_by_m1}
\end{figure}

To make this scaling even clearer, and to show its robustness over a range of system parameters, in Figs \ref{Fig:gamma_e_by_aa} - \ref{Fig:gamma_e_by_m1} we show the mean square values of $\gamma^2_e$, binned by logarithmic increments of $N_{\mathrm{ej}}/N_{\mathrm{sec}}$ for various combinations of $a_a$, $m_1$ and $m_2$. We see that the approximate scaling given by Eq. (\ref{eq:dynsec_scaling}) agrees very well with the simulations for values of $a_a/a_1$ ranging from $1/7 - 1/20$, $m_1$ from $3 - 0.3 M_J$, and $m_2/m_1$ from $1/10$ to $1/2$, although there is a trend of increasing deviation from Eq. (\ref{eq:dynsec_scaling}) when $N_{\mathrm{ej}}/N_{\mathrm{sec}} \gg 1$. We explore a possible reason for this deviation, and present a more accurate analytic formula for $\langle \gamma^2 \rangle$ in Sec.  \ref{sec:simple_model_1+2}. In general, the above scaling is accurate for $m_2/m_1 \lesssim 1/2$ and $a_a / a_{1} \lesssim 1/5$. When $m_1 \sim m_2$ and/or $a_a/a_{1,0} \gtrsim 1/5$, it is often the case that the ejected planet can come very close to the orbit of planet $a$, resulting in strong non-secular interactions that causes $\gamma_e, ~\gamma_{\theta}$ to be much greater than predicted by Eq. (\ref{eq:dynsec_scaling}). 

The simple universal scaling $\sqrt{N_{\mathrm{ej}}}$ can in fact be derived from the first principles using secular Laplace-Lagrange theory, as we discuss below.

\subsection{Analytic Model for ``1+2'' Secular Evolution: Eccentricity}
\label{sec:simple_model_1+2}
We model the dynamical evolution of an inner planet $a$ subject to the gravitational influence of a pair of outer perturbers under-going gravitational scattering as a linear stochastic differential equation (SDE). We define $\mathcal{E} \equiv e \exp{(i \varpi)}$ and $\mathcal{I} \equiv \theta \exp{(i \Omega)}$ as the complex eccentricity and inclination respectively. Note that $e = |\mathcal{E}|$ and $\theta = |\mathcal{I}|$. In the discussion below we will focus on the eccentricity evolution and derive the boost factor $\gamma_e$, although the inclination is completely analogous and will have the same scaling as $\gamma_{\theta}$.

First, consider an inner planet $m_a$ with initial eccentricity $\mathcal{E}_{a,0}$ undergoing secular evolution with an external planet $m_1 \gg m_a$ that has a constant eccentricity $\mathcal{E}_1$. For simplicity, we ignore for now the secular interaction between planet $a$ and $2$. The evolution of $\mathcal{E}_a(t)$ is governed by the ODE
\begin{equation}
    \frac{d\mathcal{E}_a(t)}{dt} = i \omega_{a1} \mathcal{E}_a(t) - i \nu_{a1} \mathcal{E}_1(t),
    \label{eq:dea_dt}
\end{equation}
where $\omega_{a1}, \nu_{a1}$ are given by Eqs. (\ref{eq:wjk})-(\ref{eq:vjk}). The solution to the above equation is given by
\begin{equation}
    \mathcal{E}_a(t) = \mathcal{E}_{a,\mathrm{free}}(t) \exp{(i\omega_{a1}t)} + \mathcal{E}_{a,\mathrm{forced}},
\end{equation}
where 
\begin{equation}
\mathcal{E}_{a,\mathrm{forced}} =  \frac{\nu_{a1}}{\omega_{a1}} \mathcal{E}_1,
\label{eq:e_force}
\end{equation}
and
\begin{equation}
\mathcal{E}_{a,\mathrm{free}} = \mathcal{E}_{a,0} - \mathcal{E}_{a,\mathrm{forced}}.
\end{equation}
Applying Eq. (\ref{eq:dea_dt}) to the secular evolution of planet $a$ after the ejection of planet $2$, we have that $\mathcal{E}_{a,0} = \mathcal{E}_{a,\mathrm{ej}}$ (where $\mathcal{E}_{a,\mathrm{ej}} = \mathcal{E}_a(t_{\mathrm{ej}})$), and the RMS eccentricity $|\mathcal{E}_{a,\infty}|$ is given by
\begin{multline}
    |\mathcal{E}_{a,\infty}|^2 = |\mathcal{E}_{a,\mathrm{free}}|^2 + |\mathcal{E}_{a,\mathrm{forced}}|^2 
    \\ =  |\mathcal{E}_{a,\mathrm{ej}}|^2 + 2|\mathcal{E}_{a,\mathrm{forced}}|^2 - 2\mathrm{Re}(\mathcal{E}_{a,\mathrm{ej}}\mathcal{E}_{a,\mathrm{forced}}^*).
    \label{eq:e_infty_sq}
\end{multline}
Note that $|\mathcal{E}_{a,\infty}|$ is what we termed $e_{a,\infty}$ in Sec. \ref{sec:1+2_empirical}. If the initial eccentricity of planet $a$ is zero, then the free eccentricity is equal to the forced eccentricity, and $e_{\mathrm{sec}} = \sqrt{2} e_{\mathrm{forced}}$.

Now we ask the question: What happens to $\mathcal{E}_a(t)$ if, instead of being a constant, $\mathcal{E}_1(t)$ is a stochastically varying quantity, as is the case during the scattering process. We study a version of Eq. (\ref{eq:dea_dt}) with $\mathcal{E}_1$ being given by a Brownian motion stochastic process: $\mathcal{E}_1(t) = Z(t)$, where $Z(t)$ is a Brownian motion in the complex plane with diffusion constant equal to $\sigma_{\mathcal{E}1}$, i.e. $Z(t) = X(t) + i Y(t)$ where $X(t), ~Y(t)$ are each given by a Gaussian distribution with mean $\langle X \rangle = \langle Y\rangle = 0$, variance $\mathrm{Var}(X) = \mathrm{Var}(Y) = \sigma^2_{\mathcal{E}1} t$, and covariance $\mathrm{Cov}[(X(s), X(t)] =\mathrm{Cov}[(Y(s), Y(t)] =  \sigma_{\mathcal{E}1}^2 \mathrm{min}(s, t)$.

The diffusion coefficient of the perturber eccentricity, $\sigma_{\mathcal{E}1}$ is a constant that can either be calculated analytically or numerically, or derived empirically from the time series of scattering planet systems. We make a heuristic estimate of it here. Over the ejection timescale, the eccentricity of planet 1 changes from $e_1 = 0 \rightarrow e_{1,\mathrm{ej}}$ (where $e_{1,\rm ej}$ is the eccentricity of planet 1 when planet 2 has been ejected; see Section \ref{sec:sec2}). On average, this process takes $N_{\rm ej} \sim t_{\mathrm{ej}}/P_{1,0} \sim b^2$ orbits (see Eq. \ref{eq:f_De}). Thus, one might surmise:
\begin{equation}
    \langle e^2_{1, \mathrm{ej}} \rangle \sim 2 \sigma_{\mathcal{E}1}^2 b^2 P_{1,0},
\end{equation}
where $\langle e^2_{1, \mathrm{ej}} \rangle \sim (m_2/m_1)$ (see Sec. \ref{sec:orbital_parameters}). This yields
\begin{equation}
    \sigma^2_{\mathcal{E}1} \sim  \langle e^2_{1, \mathrm{ej}} \rangle / (2P_{1,0} b^2).
    \label{eq:sigma_e1}
\end{equation}
We would like to know what are the mean, variance and distributions of $\mathcal{E}_a(t)$ given the initial conditions and parameters. Note that the value of $\mathcal{E}_a(t)$ at ejection is not the ultimate quantity of interest here, since planet $a$ still under-goes secular coupling with planet $1$ after ejection. Our final goal is to derive the expectation, and if possible the distribution of $\mathcal{E}_{a,\infty}$. 

To proceed, note that Eq. (\ref{eq:dea_dt}), with $\mathcal{E}_1(t) = Z(t)$, has the solution
\begin{equation}
    \mathcal{E}_a(t) = -i\nu_{a1} e^{i \omega_{a} t} \int_0^t e^{-i \omega_{a} s}   Z(s) ds,
    \label{eq:ea}
\end{equation}
where we have assumed $\mathcal{E}_a(0) = 0$. The statistical property of $\mathcal{E}_a(t)$ as determined by Eq. (\ref{eq:ea}) depends on whether the final value of $\mathcal{E}_1(t_{\mathrm{ej}}) = \mathcal{E}_{1,\mathrm{ej}}$ is known (empirically measured, or otherwise constrained by conservation laws). If $\mathcal{E}_{1,\mathrm{ej}}$ is unconstrained, then $Z(s)$ is the classic 2-D Brownian motion. If $\mathcal{E}_{1,\mathrm{ej}}$ is known {\it a priori}, then $Z(s)$ is not a Brownian motion but rather a Brownian bridge, which is given by a different density distribution that has a reduced variance towards the end of the stochastic process. We consider both cases below. In this study, since the final values of perturber properties are known, case 2 is the more appropriate one. We deal with case 1 first as a stepping stone.

\subsubsection*{Case 1: Unknown $\mathcal{E}_{1,\mathrm{ej}}$}

We study the expected value and distribution of $\mathcal{E}_a$ at the time of ejection, $\mathcal{E}_{a,\rm ej} = \mathcal{E}_{a}(t_{\mathrm{ej}})$. First, since $\langle Z(s) \rangle = 0$ for all $s$, the integral in Eq. (\ref{eq:ea}) has expectation $\langle \mathcal{E}_a (t)\rangle = 0$ for all $t$. The variance and covariances of interest can be computed using the linearity of expectation. The variance of the final eccentricity is given by (see Appendix \ref{sec:A1})
\begin{equation}
    \langle |\mathcal{E}_{a,\mathrm{ej}}|^2 \rangle = 4 \left(\frac{\nu_{a1}}{\omega_{a}} \right)^2 \left[1 - \frac{\sin{(\omega_{a1}t_{\mathrm{ej}})}}{\omega_{a1}t_{\mathrm{ej}}}\right] \sigma^2_{\mathcal{E}1} t_{\mathrm{ej}},
\end{equation}
while the covariance between the final eccentricity and its forced amount (see Eq. \ref{eq:e_force}) is 
\begin{equation}
    \langle \mathrm{Re}(\mathcal{E}_{a,\mathrm{ej}}\mathcal{E}_{a,\mathrm{forced}}^*) \rangle = 2 \left(\frac{\nu_{a1}}{\omega_{a1}} \right)^2 \left[1 - \frac{\sin{(\omega_{a1}t_{\mathrm{ej}})}}{\omega_{a1}t_{\mathrm{ej}}}\right] \sigma^2_{\mathcal{E}1} t_{\mathrm{ej}}.
\end{equation}
The expectation of the forced eccentricity is
\begin{equation}
    \langle |\mathcal{E}_{a,\mathrm{forced}}|^2 \rangle = 2 \left(\frac{\nu_{a1}}{\omega_{a}} \right)^2  \sigma^2_{\mathcal{E}1} t_{\mathrm{ej}}.
\end{equation}
From Eqs. (\ref{eq:e_infty_sq})-(\ref{eq:sigma_e1}), the RMS eccentricity of planet $a$ is 
\begin{equation}
    \langle |\mathcal{E}_{a,\infty}|^2 \rangle =
    4\left(\frac{\nu_{a1}}{\omega_{a}}\right)^2\sigma_{\mathcal{E}1}^2 t_{\mathrm{ej}} \sim \frac{25 a_a^2\langle e_{1,\mathrm{ej}} \rangle^2 N_{\mathrm{ej}}}{8 a^2_1 b^2}.
\end{equation}
We see that $\langle |\mathcal{E}_{a,\infty}|^2 \rangle \propto N_{\mathrm{ej}}$. However, in this unconstrained case, it is also the case that $|\mathcal{E}_{a,\mathrm{forced}}|^2 \propto N_{\mathrm{ej}}$, so that the scaling for the boost factor is $\gamma_e = \mathrm{const.}$, which is contrary to our empirical results. This contradiction arises because we have not taken into account the fact that $\mathcal{E}_{1,\mathrm{ej}}$ is a known quantity and not a random variable. Only when we place a constraint on the Brownian motion at $t_{\mathrm{ej}}$ can the desired scaling be derived.

\subsubsection*{Case 2: $\mathcal{E}_{1,\mathrm{ej}}$ is known or constrained}
When the final value of $\mathcal{E}_{1}$ at $t = t_{\mathrm{ej}}$ is known, the evolution $\mathcal{E}_a(t)$ is qualitatively similar, but the statistical properties change due to the Brownian motion in $\mathcal{E}_1$ being ``tied down'' at the final time, giving it a lower variance. To recognise that this process is different from an unconstrained Brownian motion, we label it $B(t)$ instead of $Z(t)$. At $t = 0$, we have $\mathcal{E}_1 = B(0) = 0$, while at $t = t_{\mathrm{ej}}$, $\mathcal{E}_1 = B(t_{\mathrm{ej}}) = \mathcal{E}_{1,\mathrm{ej}}$. In between this time, $B(t)$ executes a (complex) Brownian motion and is normally distributed, with mean and variance \citep{Borodin2002}
\begin{align}
    \langle B(t) \rangle &= \left(\frac{t }{t_{\mathrm{ej}}}\right)\mathcal{E}_{1,\mathrm{ej}}
    \label{eq:mean_B} \\ 
    \mathrm{Var}[B(t)] &\equiv \langle B^2(t) \rangle - \langle B(t) \rangle^2 =  \frac{2t(t_{\mathrm{ej}} - t)\sigma^2_{\mathcal{E}1}}{t_{\mathrm{ej}}}.
\end{align}
Another relevant quantity is the covariance of a Brownian bridge with itself at a different time, which (without loss of generality, assuming $s < t$) is given by
\begin{equation}
    \mathrm{Cov}[B(s), B(t)] \equiv \langle B(s) B^*(t) \rangle = \frac{2s(t_{\mathrm{ej}} - t)\sigma^2_{\mathcal{E}1}}{t_{\mathrm{ej}}}.
    \label{eq:covar_B}
\end{equation}
We can now calculate the expectation of $\mathcal{E}_{a,\rm ej}$. Unlike the unconstrained case, the mean is non-zero:
\begin{equation}
    \langle \mathcal{E}_{a,\mathrm{ej}} \rangle = i \mathcal{E}_{1,\mathrm{ej}} \left(\frac{\nu_{a1}}{\omega_{a1}}\right)\left(\frac{e^{i\omega_{a1}t_{\mathrm{ej}}} - i\omega_{a1}t_{\mathrm{ej}} - 1}{\omega_{a1}t_{\mathrm{ej}}} \right),
\end{equation}
and the square of the mean eccentricity is
\begin{multline}
    |\langle \mathcal{E}_{a,\mathrm{ej}} \rangle|^2 = \left(\frac{\nu_{a1}}{\omega_{a1}}\right)^2 |\mathcal{E}_{1,\mathrm{ej}}|^2 \\ \times \left[1 + 2\left(\frac{1 - \cos{(\omega_{a1} t_{\mathrm{ej}})} - \omega_{a1} t_{\mathrm{ej}} \sin(\omega_{a1} t_{\mathrm{ej}})}{\omega^2_{a1} t_{\mathrm{ej}}^2}\right) \right].
\end{multline}
The variance of the eccentricity is given by
\begin{equation}
    \langle |\mathcal{E}_{a,\mathrm{ej}}|^2 \rangle -  |\langle \mathcal{E}_{a,\mathrm{ej}} \rangle|^2 = 2\sigma^2_{\mathcal{E}1} \left(\frac{\nu_{a1}}{\omega_{a1}} \right)^2 t_{\mathrm{ej}} \left[1 - 2 \left(\frac{1 - \cos(\omega_{a1} t_{\mathrm{ej}})}{\omega^2_{a}t^2_{\mathrm{ej}}}\right) \right].
\end{equation}
In order to know the final RMS eccentricity $\mathcal{E}_{a,\infty}$, we also require the covariance between $\mathcal{E}_{a,\mathrm{ej}}$ and $\mathcal{E}_{a,\mathrm{forced}}$, which is given by
\begin{equation}
    \langle \mathrm{Re}(\mathcal{E}_{a,\mathrm{ej}} \mathcal{E}^*_{a,\mathrm{forced}})  \rangle = |\mathcal{E}_{a,\mathrm{forced}}|^2  \left[\frac{\cos{(\omega_{a1}t_{\mathrm{ej}})}-1}{\omega_{a1}t_{\mathrm{ej}}} \right].
\end{equation}
Combining these expressions with Eq. (\ref{eq:e_infty_sq}), the RMS eccentricity at infinity is given by 
\begin{multline}
    \langle |\mathcal{E}_{a,\infty}|^2 \rangle = 2\left(\frac{\nu_{a1}}{\omega_{a1}} \right)^2 \Bigg( \sigma^2_{\mathcal{E}1}  t_{\mathrm{ej}} \left[1 - 2 \left(\frac{1 - \cos(\omega_{a1} t_{\mathrm{ej}})}{\omega^2_{a1}t_{\mathrm{ej}}^2}\right) \right] 
    \\ + |\mathcal{E}_{1,\mathrm{ej}}|^2  \left[\frac{3}{2} + \frac{1 - \cos{(\omega_{a1}t_{\mathrm{ej}})} - \sin{(\omega_{a1}t_{\mathrm{ej}})}}{\omega_{a1}t_{\mathrm{ej}}} + \frac{1 - \cos{(\omega_{a1} t_{\mathrm{ej}})} }{\omega_{a1}^2 t_{\mathrm{ej}}^2} \right] \Bigg).
\end{multline}
In the above equation, when $\omega_{a1} t_{\mathrm{ej}} \ll 1$, the second term of the RHS dominates and we have $|\mathcal{E}_{a,\infty}|^2 \propto t_{\mathrm{ej}}$. On the other hand, when $\omega_{a1} t_{\mathrm{ej}} \gg 1$, the first term dominates and we also have $|\mathcal{E}_{a,\infty}|^2 \propto t_{\mathrm{ej}}$. In order words, for all $t_{\mathrm{ej}}$ we have $\langle |\mathcal{E}_{a,\infty}|^2 \rangle \propto t_{\mathrm{ej}}$, in agreement with our numerical results.
Since $e^2_{a,\mathrm{sec}} = 2|\mathcal{E}_{a,\mathrm{forced}}|^2$, the ensemble RMS of the boost factor $\langle \gamma_e^2 \rangle$ is given by
\begin{multline}
\langle \gamma_e^2 \rangle  = \frac{\langle |\mathcal{E}_{a,\infty}|^2 \rangle - 2|\mathcal{E}_{a,\mathrm{forced}}|^2}{2|\mathcal{E}_{a,\mathrm{forced}}|^2} 
\simeq Ax \left[1 - 2\left(\frac{1 - \cos{(x)}}{x^2} \right) \right]
\\ + \frac{1 - \cos{(x)} - \sin{(x)}}{x}  + \frac{1 - \cos{(x)}}{x^2} + \frac{1}{2} ,
\label{eq:langle_gamma}
\end{multline}
where we have defined $x \equiv \omega_a t_{\mathrm{ej}} \sim 2\pi N_{\mathrm{ej}} / N_{\mathrm{sec}}$, and $A$ is the dimensionless constant
\begin{equation}
    A \equiv \frac{\sigma^2_{\mathcal{E}1}}{\omega_{a1} |\mathcal{E}_{1,\mathrm{ej}}|^2} \sim \frac{1}{\omega_{a1}b^2 P_{1,0}} \sim 2\pi \left( \frac{\langle N_{\mathrm{ej}}\rangle_{\mathrm{HM}}}{N_{\mathrm{sec}}} \right),
\end{equation}
and $\langle N_{\mathrm{ej}}\rangle_{\mathrm{HM}} = b^2$ (Eq. \ref{eq:b_fit}) is the harmonic mean of $N_{\mathrm{ej}}$. Here we have made use of the fact that the final eccentricity is well constrained by conservation laws, so $\langle e_{1,\mathrm{ej}} \rangle^2 \approx |\mathcal{E}_{1,\mathrm{ej}}|^2$.

Eq. (\ref{eq:langle_gamma}) has two regimes: when $x \ll 1$, $\gamma_{e} \simeq \sqrt{x/2}$, while when $x \gg 1$, we have $\gamma_e \simeq \sqrt{Ax}$. The transition between the two regimes occurs when $x \sim \pi$. Using our earlier estimates for $b$ (Eq. \ref{eq:b_fit}) , $A$ is of order
\begin{equation}
    A \sim 7  \left(\frac{m_1}{M_{\star}} \right) \left(\frac{a_a}{a_1} \right)^{-3/2} \left(1 + \frac{m_2}{m_1} \right)^{4} \left(\frac{a_{1,0}}{a_{2,0}} \right)^{-2}.
\end{equation}
For the typical range of parameters relevant to Kepler planets ($m_1 \sim 10^{-3}$ and $a_a / a_1 \sim 1/10$) one obtains $A \sim 0.3$. Given the inherent scatter in the simulation results, the difference between the two regimes in Eq. (\ref{eq:langle_gamma}) is too subtle for us to empirically measure $A$ in this study. Thus in this paper we simply adopt the approximation $\gamma_e \sim \sqrt{N_{\mathrm{ej}}/N_{\mathrm{sec}}}$ which agrees well with the empirical results.

Having computed the mean value $\langle \gamma_e^2 \rangle$ we now comment on its distribution. The Brownian bridge has a distribution that is normally distributed over an ensemble of simulations, and any linear transformation of normally distributed variables is also normally distributed. From Eq. (\ref{eq:gamma_e}) and Eq. (\ref{eq:e_infty_sq}), the boost factor can be written as
\begin{equation}
    \gamma^2_e = \frac{ \left||\mathcal{E}_{a, \mathrm{ej}}|^2 - 2\mathrm{Re}(\mathcal{E}_{a,\mathrm{ej}}\mathcal{E}_{a,\mathrm{forced}}^*) \right|}{|\mathcal{E}_{a, \mathrm{forced}}|^{2}}.
\end{equation}
The quantities $\mathcal{E}_{a, \mathrm{ej}}$ and $\mathcal{E}_{a, \mathrm{forced}}$ are normally distributed complex variables with zero mean. In the limit that $N_{\mathrm{ej}} \gg N_{\mathrm{sec}}$, we have that $|\mathcal{E}_{a, \mathrm{ej}}|^2 \gg 2\mathrm{Re}(\mathcal{E}_{a,\mathrm{ej}}\mathcal{E}_{a,\mathrm{forced}}^*)$, and $\gamma_e$ is then the length of a 2-D vector whose components are normally distributed with zero mean; such a quantity has approximately a Rayleigh distribution. We define $\bar{\gamma}_e \equiv \langle \gamma_e^2 \rangle^{1/2}$ (see Eq. \ref{eq:langle_gamma}), then the distribution of $\gamma_e$ in this limit is given by
\begin{equation}
    f(\gamma_e) = \frac{\gamma_e}{\bar{\gamma}_e^2} \exp{\left(\frac{-\gamma^2_e}{2\bar{\gamma}_e^2}\right)}.
    \label{eq:f_gamma_e}
\end{equation}
Empirically, we find that Eq. (\ref{eq:f_gamma_e}) is a good approximation for the distribution of $\gamma_e$ even when it is not the case that $N_{\mathrm{ej}} \gg N_{\mathrm{sec}}$.

\subsection{Inclination Evolution}
\label{sec:inclination}
In the above analysis we have considered the eccentricity evolution of planet $a$ subject to a stochastic forcing by the outer perturber. The evolution of the inclination can be derived in the same manner as the eccentricity, except, whenever appropriate, replacing the complex eccentricities $\mathcal{E}$ with the corresponding complex inclinations $\mathcal{I}$, and replacing $\omega_{a1} \rightarrow -\omega_{a1}$ and $\nu_{a1} \rightarrow -\omega_{a1}$. The forced inclination is given by Eq. (\ref{eq:t_sec}). One will eventually find that the scaling for $\gamma_e$ and $\gamma_{\theta}$ is the same:
\begin{equation}
    \langle \gamma_e^2 \rangle = \langle \gamma_{\theta}^2 \rangle.
\end{equation}
In addition, the probability density distribution for $\gamma_{\theta}$ is also the same as $\gamma_e$, and is given by Eq. (\ref{eq:f_gamma_e}) (note that $\bar{\gamma}_e = \bar{\gamma}_{\theta}$). 
Since $\gamma_e, ~\gamma_{\theta}$ have the same distribution, and $\bar{\gamma}_e = \bar{\gamma}_{\theta}$, we hereafter refer to the distribution of either quantity as $\gamma$ (although note that $\gamma_e$ and $\gamma_\theta$ are uncorrelated and independently distributed).

Having computed the distribution of $\theta_{a}$, we now derive the boost factor for the mutual inclination $\gamma_{\theta,a1}$. Note that
{\color{black}
\begin{align}
    \theta^2_{a1, \infty} &= \theta^2_{a1, \mathrm{ej}} =  |\mathcal{I}_{a,\rm ej} - \mathcal{I}_{1, \rm ej}|^2 \nonumber \\
    &= |\mathcal{I}_{a,\rm ej}|^2 + |\mathcal{I}_{1, \rm ej}|^2 - 2\mathrm{Re}(\mathcal{I}_{a, \rm ej} \mathcal{I}_{1, \rm ej}^*).
\end{align}
}
From Eq. (\ref{eq:e_infty_sq}) (but replacing $\mathcal{E} \rightarrow \mathcal{I})$, we thus have
\begin{equation}
    \theta^2_{a1, \infty} = \theta^2_{a, \infty} - \theta^2_{1, \rm ej}.
\end{equation}
Recall that $\theta_{a1, \rm sec} = \theta_{a, \rm sec}/\sqrt{2}$, thus from Eq. (\ref{eq:gamma_t}) - (\ref{eq:gamma_a1}) we find
\begin{equation}
   \gamma_{\theta,a1}  = \sqrt{2} \gamma _{\theta}.
   \label{eq:two_gammas}
\end{equation}
The above equation assumes that $\theta_{a1}, ~\theta_{a} \ll 1$ and ignores the contribution from planet 2. In reality, $\gamma_{\theta,a1}$ will deviate from Eq. (\ref{eq:two_gammas}), although the above scaling still holds on average. Once we know the value of $\gamma_{\theta}$, we can convert it to the corresponding value of $\gamma_{\theta,a1}$ to obtain the mutual inclination boost factor, and vice versa.

\subsection{Marginal Distribution of the Boost Factor}
\label{sec:marginal}
The distributions we have derived so far for $\gamma_e, ~\gamma_{\theta}$ are contingent on $N_{\mathrm{ej}}$, which is not an observable quantity. However, since we have some understanding of the distribution of $N_{\mathrm{ej}}$, we can now marginalize over it and only deal with observable quantities.  First, combining Eq. (\ref{eq:Nej_dist}) and Eq. (\ref{eq:f_gamma_e}) we can write the joint distribution for $N_{\mathrm{ej}}$ and $\gamma$ as
\begin{equation}
    f(N_{\mathrm{ej}}, \gamma) = \frac{b \gamma}{\bar{\gamma}^2 \sqrt{2\pi N_{\mathrm{ej}}^3}} \exp{\left(\frac{-b^2}{2N_{\mathrm{ej}}}\right)} \exp{\left(\frac{-\gamma^2}{2\bar{\gamma}^2}\right)}.
    \label{eq:fjoint}
\end{equation}
Now, from Eq. (\ref{eq:dynsec_scaling}) we have that $ \bar{\gamma}^2 \sim N_\mathrm{ej}/N_{\mathrm{sec}}$. Substituting into Eq. (\ref{eq:fjoint}), and integrating over $N_{\mathrm{ej}}$ we thus obtain the distribution for $\gamma$ in terms of observable quantities only:
\begin{align}
    f(\gamma) &= \int_0^{\infty} b\gamma N_{\mathrm{sec}}  \sqrt{\frac{1}{2\pi {N_{\mathrm{ej}}^5}}}     \exp{\left(\frac{-b^2 -\gamma^2 N_{\mathrm{sec}} }{2N_{\mathrm{ej}}}\right)} ~ d N_{\mathrm{ej}}  \nonumber
    \\ &= \frac{b \gamma N_{\mathrm{sec}} }{(b^2 + N_{\mathrm{sec}} \gamma^2)^{3/2}}.
\end{align}
{\color{black}
Now, we define $y$ as the `normalized' boost factor
\begin{equation}
    y \equiv \gamma \sqrt{N_{\mathrm{sec}}/\langle N_{\mathrm{ej}}\rangle_\mathrm{HM}},
    \label{eq:y_definition}
\end{equation}
(recall that $b^2 = \langle N_{\rm ej} \rangle_{\rm HM}$), then we have the rather elegant expression for the normalized boost factor $y$:
\begin{equation}
     f(y) = \frac{y}{(1 + y^2)^{3/2}}.
     \label{eq:f_y}
\end{equation}
}
In the distribution above, the probability that $y$ is greater than some constant $y'$ is given by
\begin{equation}
    P(y \ge y') = \frac{1}{\sqrt{1 + y'^2}}.
    \label{eq:f_y_CDF}
\end{equation}
Just like the distribution for $N_{\mathrm{ej}}$ (Eq. \ref{eq:Nej_dist}), the distribution $f(y)$ is a long-tailed one, such that all its higher moments (e.g. mean, variance) fail to exist. Its mode occurs at $y = 1/\sqrt{2}$, its geometric mean is $\langle y \rangle_{\mathrm{GM}} = 2$, its harmonic mean is $\langle y \rangle_{\mathrm{HM}} = 1$ and its median is $y = \sqrt{3}$. The 68\% and 95\% confidence intervals are $y \in [0.65, 6.2]$ and $y \in [0.23, 40]$ respectively. Assuming that $a_{2,0} \sim a_{1,0}$, the harmonic mean of $\gamma$ is given by the following scaling:
\begin{equation}
    \langle \gamma \rangle_{\mathrm{HM}} = \sqrt{\langle N_{\mathrm{ej}}\rangle_\mathrm{HM}/N_{\mathrm{sec}}} \sim 1.1  \left( \frac{m_1}{M_{\star}}\right)^{-1/2} \left( \frac{a_a}{a_{1,0}}\right)^{3/4} \left(1+ \frac{m_2}{m_1}\right)^{2}.
    \label{eq:y_hm}
\end{equation}
{\color{black} From this scaling, we see that $\langle \gamma \rangle_{\mathrm{HM}} = y/\gamma$, thus we re-interpret $y = \gamma / \langle \gamma \rangle_{\mathrm{HM}}$ as the boost factor `normalized' by its harmonic mean and define 
$y_e = \gamma_e/\langle \gamma_e \rangle_{\mathrm{HM}}, ~y_\theta = \gamma_\theta/\langle \gamma_\theta \rangle_{\mathrm{HM}}$ for the normalized eccentricity and inclination boost factors respectively. Note that the scaling relation in Eq. (\ref{eq:y_hm}) applies equally to $y_e$ and $y_{\theta}$. We see that the effect of CJ scatterings on inner planets is the greatest if the CJ scatters are lower in mass, have semi-major axes more comparable to the inner planets, and have comparable masses.}

\begin{figure}
\includegraphics[width=1.01\linewidth]{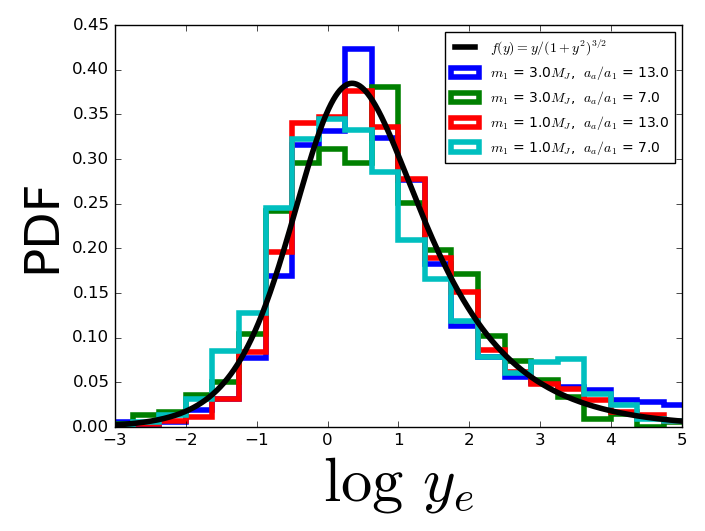}
\includegraphics[width=1.01\linewidth]{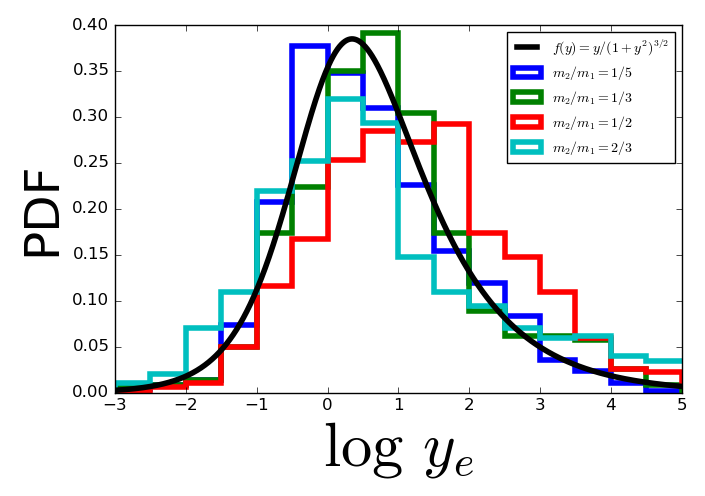}
\caption{Distribution of $y_e \equiv  \gamma_e  \sqrt{N_{\mathrm{sec}}/\langle N_{\mathrm{ej}}\rangle_\mathrm{HM}}$ (see Sec. \ref{sec:marginal}). The histograms are empirical distributions obtained from our simulations, while the black line is the theoretical distribution given by Eq. (\ref{eq:f_y}) - (\ref{eq:y_hm}). On the top panel, $m_2/m_1 = 1/5$ while $m_1$ and $a_a/a_1$ varies as shown in the legend. On the bottom panel, $a_a/a_1 = 1/10$ and $m_1 = 1 M_J$, while $m_2/m_1$ varies as shown in the legend. }
\label{fig:gamma}
\end{figure}

In Fig. \ref{fig:gamma} we show a comparison between our theoretical distribution given by Eq. (\ref{eq:f_y}) for the normalized eccentricity boost factor $y_e$ and the empirical distribution from our suite of simulations. We find that for $m_2/m_1 \lesssim 1/3$, the theoretical distribution agrees well with the empirical one over a range of different masses and $a_a/a_1$. The empirical distribution starts to deviate somewhat from Eq. (\ref{eq:f_y}) for more comparable masses: in particular, the distribution becomes even more heavy-tailed, with significant fraction having $y \gg 1$, although the empirical mode and harmonic mean still agreed with Eq. (\ref{eq:y_hm}) to with-in a factor of a few.

\subsection{Theoretical Model: Simplifications and Refinements}
In developing our stochastic model for ``1+2'' scattering, we have made several simplifying assumptions. A more careful treatment can yield refinements to the model and more accurate estimates for the distribution of final parameters. We discuss the most crucial simplifications and suggest possible ideas for refinement below.
\begin{itemize}
    \item {\bf Secular forcing by planet 2:} In our theoretical model we have ignored the secular interaction between the inner planet and planet $2$ as it is being ejected from the system. This can be justified in the limit that $m_2/m_1 \ll 1$. However, for more comparable masses, $m_2$ can have an equal or even greater effect than $m_1$ on the secular evolution of the inner system. Our simplification of ignoring planet 2 is the main reason why our estimate from Eq. (\ref{eq:langle_gamma}) becomes less accurate when $m_2 \sim m_1$. Since at the end of the ejection process, the secular forcing by $m_2$ vanishes, one way to incorporate the influence of $m_2$ is to absorb it into the variance of the Brownian bridge, i.e. by replacing $\sigma_{\mathcal{E}1} \rightarrow \sigma_{\mathcal{E}1} (1 + \kappa_{12})$, where $\kappa_{12}$ is a dimensionless ratio that depends on $m_2/m_1$ (and possibly other quantities) that accounts for the added effect of secular perturbations by $m_2$. For certain initial configurations, 3-body secular interactions can also give rise to secular resonances that would increase the amount of eccentricity and inclination excited in the inner planet \citep[see][]{Lai2017,Pu2018}.
    \item {\bf Linearity in $\mathcal{E}, ~\mathcal{I}$:} In our theoretical model we have assumed that the secular evolution in eccentricity and inclination is linear. Note however that our hybrid algorithm (Sec.  \ref{sec:hybrid_algo}) allows for the possibility of larger growths in eccentricity due to non-linear Lidov-Kozai oscillations, and that such oscillations are indeed possible when $\theta_{a}$ grows to large values. Unfortunately differential equations with such stochastic terms become intractable when stochasticity is involved, and one would have to resort to numerical integrations in this regime.
    \item {\bf Constancy of $a_1$:} In our theoretical model we have also assumed that $a_1$ (and therefore $\omega_{a1}, ~\nu_{a1}$) is constant, which is approximately the case when $m_2 \ll m_1$ but breaks down at more comparable mass ratios. In reality, $a_1$ changes randomly as $a_2$ undergoes strong scatterings, and its final value can decrease by as much as $a_{1,\mathrm{ej}}/a_{1,0} = 1/2$ in the limit that $m_2 = m_1$. There are two ways to refine our model to incorporate this: First, one can absorb the stochastic changes in $\nu_{a1}$ as additional variance in $\sigma_{\mathcal{E}1}$, i.e. by replacing $\sigma_{\mathcal{E}1} \rightarrow \sqrt{\sigma^2_{\mathcal{E}1} + \sigma^2_{\nu 1}}$, where $\sigma^2_{\nu 1}$ is the RMS change in $\nu_{a1}$ per unit time. In addition, one should replace $\omega_a$ with its expectation, i.e.
    \begin{equation}
        \langle \omega_a(t) \rangle = \omega_{a,0} + (\omega_{a,\mathrm{ej}} - \omega_{a,0})(t/t_{\mathrm{ej}}).
    \end{equation}
    The above addition still allows for an analytic estimate for the final eccentricity and inclination, while incorporating the non-constancy of $a_1$, although the resulting final expressions are much less elegant.
    \item  {\bf Flat power spectrum of $\sigma_{\mathcal{E}1}$}: We assume that the $\sigma_{\mathcal{E}1}$ is a constant that is independent of timescale. In reality, this assumption could break down at timescales much shorter than the orbital timescale of the outer giant planets, and the scaling $\gamma \propto \sqrt{N_\mathrm{ej}/N_\mathrm{sec}}$ would break down. This would be most pertinent in cases where $\omega_a \gg 1/P_{1,0}$, and would lead to an over-estimation of the boost factors.
\end{itemize}

\section{Extension to More Inner Planets}
\label{sec:2+2}
Having understood the dynamics of ``1+2'' scattering we now generalize our results to the case with more than one inner planets. The parameter space is vast when additional planets are considered, but as we shall demonstrate, the universal scalings given by Eqs. (\ref{eq:dynsec_scaling}) and (\ref{eq:f_y}) - (\ref{eq:y_hm}) remain valid.

\subsection{Two inner planets}

For each of our N-body simulations, we consider inner systems with $a_a = a_1/20$ and $a_b = 1.5 a_a$, and $m_a = m_b = 3 M_{\oplus}$. The initial eccentricities and inclinations of the inner planets are set to zero. In our simulations, the inner planets effect each other secularly, and are influenced by the outer perturbers through secular interactions, as described by Sec.  \ref{sec:hybrid_algo}. 

For systems with 2 inner planets and an external perturber, the dynamics of the system depends crucially on the dimensionless coupling parameter $\epsilon_{ab}$ \citep{Lai2017, Pu2018}, given by
\begin{equation}
    \epsilon_{ab} \equiv \frac{\omega_{b1} - \omega_{a1}}{\omega_{ab} + \omega_{ba}} \approx \left(\frac{m_1}{m_b} \right) 
     \left(\frac{a_b}{a_1} \right)^{3}
      \left[\frac{3 a_a/a_b}{b_{3/2}^{(1)}(a_a/a_b)} \right]
       \frac{(a_b/a_a)^{3/2} - 1}{1 + (L_a/L_b)},
\end{equation}
where $L_i \equiv m_i \sqrt{GM_{\star}a_i}$ is the circular angular momentum of the planet, and $b_{3/2}^{(1)}(a_a/a_b)$ is the Laplace coefficient given by Eq. (\ref{eq:laplace_coef}).

\begin{figure}
\includegraphics[width=0.95\linewidth]{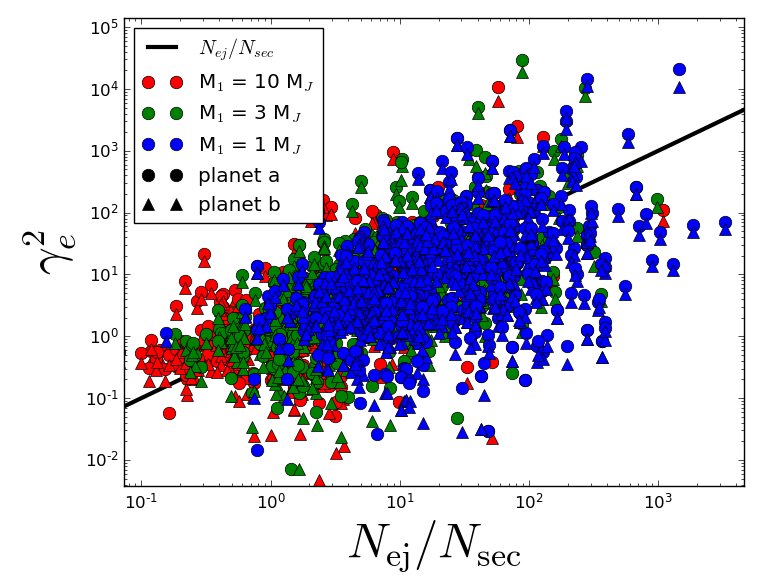}
\caption{Similar to Fig. \ref{Fig:Fig8}, except with 2 inner planets. We have $m_a = m_b = 3 M_{\oplus}$, $a_a = a_1/20$ and $a_b = 1.5 a_a$, while $m_1$ varies as shown on the plot legend and $m_2 = m_1/5$. The boost factor for the first inner planet $\gamma_{e,a}$ corresponds to the filled circles, while that for the second inner planet is shown as filled triangles. }
\label{Fig:4pl_a}
\end{figure}

\begin{figure}
\includegraphics[width=1.0\linewidth]{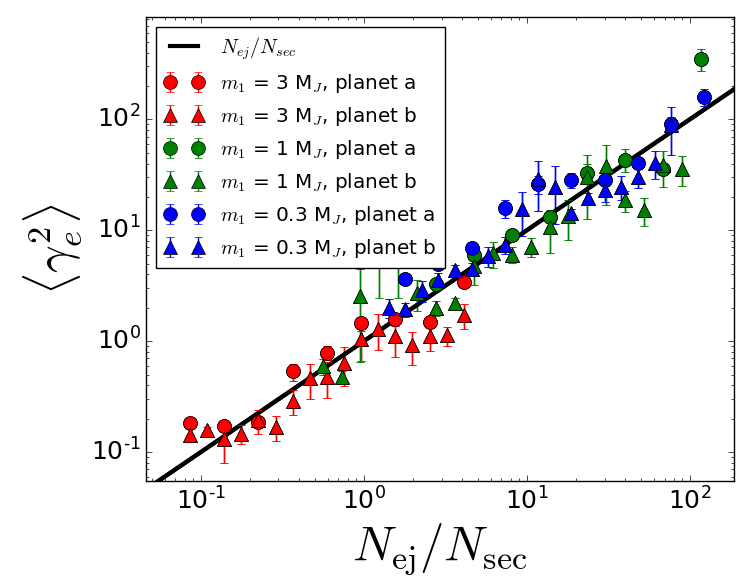}
\includegraphics[width=1.0\linewidth]{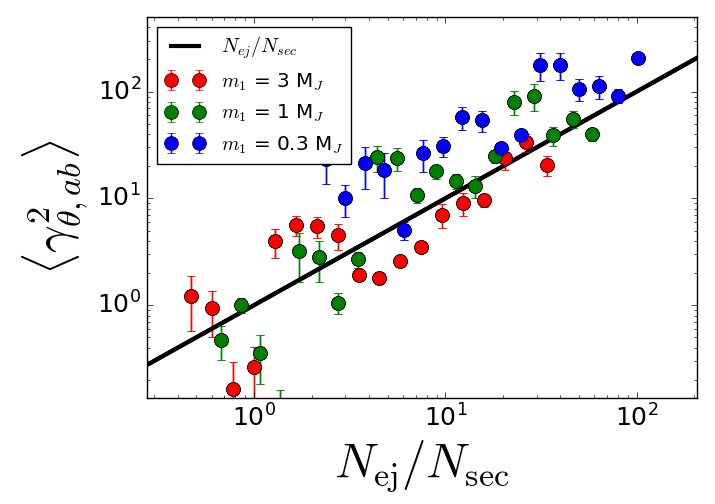}
\caption{Similar to Fig. \ref{Fig:gamma_e_by_m1}, except the simulations have two inner planets. The system parameters are the same as those for Fig. \ref{Fig:4pl_a}. The top panel shows the eccentricity boost factor $\gamma_e^2$ while the bottom panel show the mutual inclination boost factor $\gamma_{\theta,ab}^2$. }
\label{Fig:4pl_e_th}
\end{figure}

In the parameter regime that we study in this work, the two inner planets are invariably in the ``strong coupling'' regime ($\epsilon_{ab} \ll 1$). In this limit, assuming initially circular and co-planar orbits for planets $a$ and $b$, the ``secular'' eccentricities and mutual inclinations are given by \cite[see][]{Pu2018}
\begin{align}
   e_{a,\mathrm{sec}} &= \sqrt{2} \left(\frac{\nu_{a1}\omega_b + \nu_{ab}\nu_{b1}} {\omega_a \omega_b - \nu_{ab}\nu_{ba}}\right) e_{1,\mathrm{ej}} ,\\
   e_{b,\mathrm{sec}} &= \sqrt{2} \left(\frac{\nu_{b1}\omega_a + \nu_{ba}\nu_{a1}} {\omega_a \omega_b - \nu_{12}\nu_{21}}\right) e_{1,\mathrm{ej}}  , \\
   \theta_{a1, \mathrm{sec}} &= \theta_{b1, \mathrm{sec}} \approx \theta_{1,\mathrm{ej}},  \\
    \theta_{ab,\mathrm{sec}} &=  2  \left( \frac{\omega_{a1} - \omega_{b1}}{\sqrt{(\omega_a - \omega_b)^2 + 4\omega_{ab}\omega_{ba}}} \right) \theta_{1,\mathrm{ej}},
\end{align}
where $\omega_{a} = \omega_{ab} + \omega_{a1}$ and $\omega_{b} = \omega_{ba} + \omega_{b1}$ respectively. From these ``secular'' values, we compute the values of $\gamma_{e,a}, ~\gamma_{e,b}$ and $\gamma_{\theta,ab}$ analogous to Sec.  \ref{sec:1+2_empirical}. We show the results of our simulations in Figs. \ref{Fig:4pl_a} - \ref{Fig:4pl_e_th}. We see that in the ``2+2'' case the boost factor is still consistent with the scaling law Eq. (\ref{eq:dynsec_scaling}), even though the values of $\omega_a$, $\omega_b$ and the forced eccentricities and inclinations are given by very different expressions. 

\subsection{3 or More Inner Planets}
\label{sec:3+2}
Having briefly studied the ``2+2'' scattering we make some remarks on extending our theory to systems with 3 or more inner planets. The numerical algorithm described in Sec.  \ref{sec:hybrid_algo} works for a general number of inner (and outer) planets, so long as the inner and outer systems are sufficiently detached that the outer planets do not come in close contact with the inner planets. However, the theoretical model in Sec.  \ref{sec:simple_model_1+2}, and in particular Eq. (\ref{eq:langle_gamma}) must be modified if there are additional of more inner planets, due to the more complex secular coupling between the inner planets. In particular, one should deal with the amplitudes of the planet eccentricity and inclination secular eigenmodes, and the secular precession frequency should be replaced with the mode frequencies. The (complex) eigenmode amplitude of the $\alpha$-th mode should scale as 
\begin{equation}
\mathcal{E}_{\alpha,\mathrm{ej}} \propto \mathcal{I}_{\alpha,\mathrm{ej}} \propto \sqrt{N_{\mathrm{ej}}/N_{\mathrm{\alpha,sec}}},
\end{equation}
where $\mathcal{E}_{\alpha,\mathrm{ej}}$, $\mathcal{I}_{\alpha,\mathrm{ej}}$ are the complex amplitude of the $\alpha$-th eccentricity and inclination eigenmodes respectively, and 
\begin{equation}
    N_{\mathrm{\alpha, sec}} \equiv \left(\frac{\omega_{\alpha,0}P_{1,0}}{ 2\pi} \right)^{-1},
\end{equation}
where $\omega_{\alpha,0}$ is the initial eigenfrequency of the $\alpha$-th eigenmode. An empirical test of the above scaling is beyond the scope of this work, but is promising ground for further research.

\section{Summary and Discussion}
\label{sec:discussion}

\subsection{Summary}
In this work we have studied CJ scatterings and their effect on inner planet systems. Our main results are summarized below.
\begin{itemize}
    \item {\bf Final outcome of CJ scattering:} We have re-examined final outcomes of strong scatterings between two CJs on gravitationally unstable orbits. At the semi-major axis of a few au or larger, the most likely outcome of such scatterings is ejection of the less massive planet \citep[see also][]{li2020}. The remaining planet, which we call planet 1, has a final semi-major axis that is consistent with orbital energy conservation. The final eccentricity and inclination of the planet is  $e_{1,\mathrm{ej}} \sim 0.7m_2/m_1$ and $\theta_{1,\mathrm{ej}} \sim 0.7\theta_{2,0}m_2/m_1$ for $m_2/m_1 \lesssim 0.5$, where $m_2$ is the mass of the ejected planet and $\theta_{2,0}$ is the initial mutual inclination of the two planets.
    \item {\bf Ejection timescale:} The timescale from the first planet-planet Hill sphere crossing to the final ejection of planet $2$ can be understood as the stopping time of a Brownian motion. We empirically measure the normalized dimensionless RMS energy exchange ($|\delta E_{12}/E_{2,0}|$) per pericenter passage $b$ over an ensemble of N-body simulations, and present a best-fit law for it in Eq. (\ref{eq:b_fit}). Given $b$, the distribution of $N_{\mathrm{ej}}$ (the number of orbits of $m_2$ prior to ejection) agrees well with Eq. (\ref{eq:Nej_dist}). 
    \item {\bf Minimum $a_2$ of ejected planet:} We find that the possible values of $a_2$ during the strong scattering and ejection is constrained by energy conservation, angular momentum conservation, and the requirement that the system cannot spontaneously scatter itself into an indefinitely stable state. Fig. \ref{Fig:inward_a} shows our empirical results for the minimum value of $a_2$ and $r_2$ over the course of ejection. We find that generally, $a_{2,\mathrm{min}} \sim a_{1,0}/2$, and for $m_2/m_1 \ll 1$ we have $r_{2,\mathrm{min}} \sim a_{1,0}/4$, although $r_{2,\mathrm{min}}$ decreases strongly as $m_2/m_1$ increases. 
    \item {\bf ``1+2'' Scattering - Numerical Results:} For well-separated inner super-Earth and outer CJ systems, the effect of CJ scatterings on the inner planet is secular. We develop a hybrid algorithm to simulate such systems efficiently, by computing two CJ scatterings and then simulating their effects on the inner planet via secular evolution. We have performed such numerical integrations for ``1+2'' systems over a wide range of parameters. We find that the eccentricity and inclination of the inner planet induced by CJ scatterings can be much larger than the secular values (Eqs. \ref{eq:e_sec} - \ref{eq:t_sec}) generated by the remaining giant planet, and the enhancement increases with $N_{\rm ej}$ (see Figs. \ref{Fig:Fig11} - \ref{Fig:Fig12}). Despite the diversity of initial parameters and final outcomes, the dynamics of the system can be succinctly summarized by the dimensionless ``boost'' factor $\gamma$ (Eqs. \ref{eq:gamma_e} - \ref{eq:gamma_t}). In the range of parameters we considered we find that Eq. (\ref{eq:dynsec_scaling}) provides a universal scaling law for the final eccentricity and inclination of the inner planet, as a function of the system parameters (see Figs. \ref{Fig:Fig8} - \ref{Fig:gamma_e_by_m1}).
    \item {\bf ``1+2'' scattering - Theoretical model:} We develop a theoretical model to explain the empirical scaling law in Eq. (\ref{eq:dynsec_scaling}), by modelling the ``1+2'' scattering process as a linear stochastic differential equation. We compute analytically the expected moments and distributions for the final inner planet eccentricity and inclination in terms of the boost factors, which are given by Eqs. (\ref{eq:langle_gamma}) - (\ref{eq:f_gamma_e}). We calculate the distribution of $\gamma$, averaged over all possible $N_{\mathrm{ej}}$, to derive a universal distribution function for the boost factor in terms of observable quantities only (Eq. \ref{eq:f_y}); this analytical distribution agrees well with empirical results (see Fig. \ref{fig:gamma}).
    \item {\bf Extension to ``2+2'' systems:} We have extended our empirical investigation to ``2+2'' systems. We find that analogous to ``1+2'' systems, Eq. (\ref{eq:dynsec_scaling}) is still valid for describing the dynamics of the system, although the final values of eccentricities and inclinations are substantially different due to strong secular coupling between the inner planets. We also describe how the theoretical model in Sec.  \ref{sec:simple_model_1+2} can be extended to inner systems with 3 or more planets.
\end{itemize}

\subsection{Caveats}
\label{sec:caveats}
In our analysis we have considered the ``clean'' cases. Several important physical effects were neglected, and we comment on them below.
\begin{itemize}
\item  {\bf Direct scatterings between the inner planet and outer giants}: In this model we have ignored the possibility of direct hard scattering between the inner planet system and the outer giants. In our simulations, cases where the inner planet crosses orbits with one of the outer giants is discarded from our tabulated results. As we have discussed in Sec. \ref{sec:inwards}, it is possible albeit unlikely for one of the giant planets to meander deeply inwards during the scattering process. For $m_1 \gtrsim 3M_J$ and $m_2 \simeq m_1$, we expect such orbit crossings to occur a small fraction ($\sim 20\%$) of the time for $a_a/a_{1,0} = 1/10$, while inner planets with $a_a/a_{1,0} \lesssim 1/20$ are generally protected from participating directly in scatterings with the giant planets. Since direct scatterings between in the inner planet and outer giants can lead to even greater excitation in eccentricity and inclination, our model thus under-estimates the potential to excite large eccentricities and inclinations in the inner planet during ``1+2'' scattering. 
\item {\bf Physical collisions between CJs: } We have focused on scatterings between CJs that result in ejection of the less massive planet. A small fraction of systems will under-go collisional mergers instead. If the final values of $e_1, ~\theta_1$ are known, then our theoretical model in Sec.  \ref{sec:simple_model_1+2} applies equally to systems that result in collisions. However, the collisional case is less interesting in terms of its impact on the inner planetary system, because the collisional timescale tends to be much shorter due to collisional probability being highest at the initial time when planet eccentricities are low \citep{Nakazawa1989, Ida1989}. In addition, the final eccentricity $e_1$ and inclination $\theta_1$ of the merger product tend to be low, due to collisions between CJs being highly inelastic \citep[see][]{li2020}. Typically, one can assume that the scattering history is unimportant for systems that result in collisions (i.e. the boost factor $\gamma \ll 1$).
\item {\bf Spin-orbit coupling:} We have neglected the coupling between the planets and stellar spin. In reality, the stellar spin and the inner planets can exchange angular momentum, which can change the inclination of the inner planets. Incorporating such evolution into our theoretical model is beyond the scope of this work. In terms of inclination evolution, including spin-orbit coupling is equivalent to adding an extra inner planet \citep[see][]{Lai2018}.
\item {\bf Short-ranged forces:}
In this study we assumed that the inner planets are effected by secular forces from other planets only. In particular, we have ignored the effects of short-ranged forces, such as general relativistic (GR) apsidal precession, tidal precession, and tidal dissipation \cite[a discussion for the relative importance of these effects is given in][]{Pu2019}. The most important such effect is GR apsidal precession, whose angular frequency (in the limit that $e_j \ll 1)$
\begin{equation}
    \omega_{j,\mathrm{GR}} = \frac{3GM_{\star}}{c^2 a_j} n_j \approx 6\times 10^{-6} \left(\frac{M_{\star}}{M_{\odot}} \right)^{3/2} \left(\frac{a_j}{0.1 \mathrm{au}} \right)^{-5/2} \mathrm{yr}^{-1}.
\end{equation}
The main effect of this additional precession is to suppress eccentricity generation. We define $\epsilon_{j1,\mathrm{GR}}$ as the ratio between $\omega_{j,\rm GR}$ and the apisdal precession frequency due to secular coupling (between planets $j$ and $1$):
\begin{equation}
    \epsilon_{j1,\mathrm{GR}} \equiv \frac{\omega_{j,\mathrm{GR}}}{\omega_{j1}} = \frac{3GM_{\star}^2 a_1^3}{a_j^4 c^2 m_1}.
\end{equation}
In the ``1+2'' case, the secular frequency of planet $a$ is thus changed from $\omega_{a1}$ to 
\begin{equation}
    \omega_a = \omega_{a1} ( 1 + \epsilon_{a1,\mathrm{GR}}),
\end{equation}
and the mean eccentricity boost factor from (Eq. \ref{eq:y_hm}) becomes
\begin{equation}
    \langle \gamma_e \rangle_{\rm HM} \sim 1.1  \left( \frac{m_1}{M_{\star}}\right)^{-1/2} \left( \frac{a_a}{a_{1,0}}\right)^{3/4} \left(1+ \frac{m_2}{m_1}\right)^{2} (1 + \epsilon_{a1, \rm GR})^{1/2}.
    \label{eq:y_hm_gr}
\end{equation}
Note that the above equation applies only to $\langle \gamma_e \rangle_{\rm HM}$ and not the inclination. Now the forced eccentricity on planet $a$ is proportional to $e_{a,\mathrm{forced}} \propto (1 + \epsilon_{a1,\mathrm{GR}})^{-1}$, at the same time we also have $\langle \gamma_e \rangle \propto (1 + \epsilon_{a1,\mathrm{GR}})^{1/2}$, thus the final eccentricity raised on planet $a$ after scattering scales as $e_{a,\infty} \propto (1 + \epsilon_{a1,\mathrm{GR}})^{-1/2}$. 

In comparison, in the purely ``secular'' scenario without scattering events, the final eccentricity raised is proportional to $e_{a,\mathrm{forced}} \propto (1 + \epsilon_{a1,\mathrm{GR}})^{-1}$. Thus we see that in the stochastic forcing case, short ranged forces such as GR apsidal precession still suppresses eccentricity generation, but the suppression factor is only proportional to the inverse square root of the strength of the short-ranged force. 

\end{itemize}

\subsection{Application to Specific Systems}

We discuss our results in the context of a few specific planet systems of interest. These systems feature an inner planet well separated from an exterior CJ with high orbital eccentricities and/or mutual inclinations. Such eccentric CJs are a natural consequence of strong scatterings between CJs. As disussed below, the observed orbital properties of these inner-outer systems can be explained using our model.
\begin{itemize}
    \item {\bf HAT-P-11} is a system with a transiting inner mini-Neptune (HAT-P-11b, $m_a = 23.4\pm 1.5 M_{\oplus}, ~a_a = 0.0525\pm0.0007$ au.) first discovered by photometry \citep{Bakos2010} and an outer CJ (HAT-P-11c) with $m_1 \sin{I_1} = 1.6\pm0.1 M_J$ and $a_1 = 4.13 \pm 0.3$ au around a mid-K dwarf with $M_{\star} = 0.81 M_{\cdot}$. RV measurements report values of $e_a = 0.218 \pm 0.03$ and $e_1 = 0.6 \pm 0.03$ for the two planets. The orbit of HAT-P-11c is highly misaligned relative to the stellar spin $\lambda_{a} \sim 100$ deg \citep{Winn2010}. \cite{Yee2018} argued that such a misalignment can be explained if the two planets are also highly mutually inclined with $\theta_{a} \gtrsim 50$ deg. {\color{black}This argument is supported by recent measurements by  \citep{Xuan2020}, who found that $54^{\circ} < \theta_{bc} < 126^{\circ}$ at the $1\sigma$ level.}
    
    Due to the very tight orbit of HAT-P-11b, GR apsidal precession is important, with $\epsilon_{a1,\mathrm{GR}} \approx 133$. Note that despite the large inclination between HAT-P-11b and HAT-P-11c, Kozai-Lidov oscillations are suppressed due to the strong GR effect, and the forced eccentricity is very small ($e_{a,\mathrm{forced}} \sim 1.1\times10^{-4}$), and the required eccentricity boost factor is $\gamma_e \sim 2000$. The observed value of $e_a$ is thus highly incompatible with pure secular interactions without scattering history.
    
    Since $e_1 = 0.6$, if the observed eccentricity is the result of strong scattering between HAT-P-11c and an ejected planet, it is most likely that $m_2 \sim m_1$ (see Sec.  \ref{sec:sec2}). Thus, applying Eq. (\ref{eq:y_hm_gr}) we have $\langle \gamma_e \rangle_{\rm HM} \sim 40$. The observed value of $\gamma_e$ is therefore larger than its typical value by a factor of $y_e = \gamma_e / \langle \gamma_e \rangle_{\mathrm{HM}} \sim 50$. According Eqs. (\ref{eq:y_definition}) - (\ref{eq:f_y_CDF}), the likelihood of seeing such a boost factor is $P(y_e \ge 50) = 0.02$. However, Eq. (\ref{eq:f_y_CDF}) underestimates $y_e$ at larger values when $m_2 \sim m_1$ (see Fig. \ref{fig:gamma}); from our empirical results we find that for $m_2/m_1 \gtrsim 0.7$, $P(y_e \ge 50) \sim 0.09$. In other words, there is a $9\%$ chance to have $e_a \gtrsim 0.2$ as a result of ``1+2'' scattering as given by the currently observed parameters.
    
    Now turning to the mutual inclination, since the nodal precession is not affected by GR precession, we have $\langle \gamma_{\theta} \sim 3.5$ (Eq. \ref{eq:y_hm}). On the other hand, the `forced' mutual inclination depends on $\theta_{12,0}$, the initial misalignment angle between HAT-P-11c and the ejected planet. The actual value of $y_\theta$ is given by $y_\theta = \theta_{a}/(3.5 \sqrt{2}\theta_{12,0}) - 1$ (recall that the factor $\sqrt{2}$ arises due to the boost factor being larger for the mutual inclination; see Sec. \ref{sec:inclination}). If we take $\theta_{a} = 50$ deg. and $\theta_{12,0} = 3$ deg., then $y_\theta \sim 3$ and $P(y_\theta \ge 3) \sim 0.4$, i.e. there is a 40\% chance for the observed mutual inclination to be as large as $50$ degrees. The probability decreases if $\theta_{12,0}$ is smaller: for $\theta_{12,0} = 1$ deg., the p-value decreases to $P(y_\theta \ge 9) \sim 0.1$. Note again that the empirical value of $P$ is greater than predicted by Eq. (\ref{eq:f_y_CDF}) due to the fact that $m_1 \sim m_2$.
    
    We conclude that for the HAT-P-11 system, the observed eccentricity of the inner planet is marginally consistent with ``1+2'' scattering with a p-value of $P \sim 0.1$ for the observed eccentricity boost factor, while the observed inclination is consistent with ``1+2'' scattering (at $P = 0.1$ level) for $\theta_{12,0} \gtrsim 1$ degree.
    
    \item{\bf Gliese 777 A} is a two-planet system detected by RV with an inner planet with $m_a \sin_{I_a} = 18 \pm 2 M_{\oplus}$ and $a_a = 0.13 \pm 0.008$ au., and an outer CJ with $m_1 \sin{I_1} = 1.56\pm 0.13 M_J$ and $a_1 = 4 \pm 0.2$ au, orbiting around a yellow subgiant with $M_{\star} = 0.82 \pm 0.17 M_{\cdot}$ \citep{Write2009}. RV measurements report $e_a \approx 0.24 \pm 0.08$ and $e_1 \approx 0.31 \pm 0.02$. 
    
    The value of $\epsilon_{a1,\mathrm{GR}} \sim 3$ which gives a forced eccentricity of $3.5\times10^{-3}$ and boost factor $\gamma_e \sim 67$, thus the value of $e_a$ cannot be explained by pure secular forcing alone. Hypothesizing that the current value of $e_1$ is due to scattering with an ejected planet, the value of $e_1 \approx 0.3$ suggests that $m_2/m_1 \sim 0.4$, which gives $\langle \gamma_e \rangle_{\rm HM} \sim 8$ and $y_e \sim 8$. Evaluating Eq. (\ref{eq:f_y_CDF}), we find that $P(y_e \ge 8) \approx 0.12$. Thus, even though the observed value of $e_a$ is much greater than the amount predicted by pure secular forcing, it is still consistent with ``1+2'' scattering theory.
    
    \item{\bf $\pi$ Men} is a two-planet system with an inner transiting super-Earth ($m_a = 4.8 M_{\oplus}$, $a_a = 0.0684$ au) discovered by TESS \citep{Huang2018TESS} and an external companion discovered by RV with $a_1 = 3.3$ au and $m_1 \approx 12.9 M_J$. The host-star is G type with $M_\star = 1.11 M_{\odot}$. Follow-up surveys have shown a significant orbital misalignment between $m_1$ and $m_a$, with $49 ~\mathrm{deg.} < \theta_{a1} < 131 ~\mathrm{deg.}$ at $1 \sigma$ level (\citealt{Xuan2020}; see also \citealt{Damasso2020, rosa2020}). The external companion has an eccentric orbit of $e_1 \approx 0.642$ while the inner planet has $e_a \approx 0.15$ \citep{Damasso2020}. 
    
    For this system $\epsilon_{a1, \rm GR} = 1.21$, and $e_{a,\rm forced} = 0.013$, thus $\gamma_e \approx 11$, which shows the current value of $e_a$ is inconsistent with pure secular forcing from $m_1$ alone. If the current value of $e_1$ is due to strong scattering, the ejected planet likely has $m_2 \sim m_1$, corresponding to $\langle \gamma_e \rangle_{\rm HM} \sim 3.3$ when GR precession is taken into account. Thus $y_e \sim 3$, which is consistent with ``1+2'' scattering with $p(y_e \ge 3) \sim 0.3$. Thus we conclude that the observed value of $e_1$ is highly compatible with ``1+2'' scattering.
    
    Now turning to the mutual inclination, we have that $\langle \gamma_\theta \rangle_{\rm HM} \sim 2.3$. Taking a fiducial value of $\theta_{a1} \approx 90$ deg., we have $y_\theta = 90 ~\mathrm{deg.}/(2.3 \sqrt{2} \theta_{12,0}) - 1$. If $\theta_{12,0} = 3$ deg., then $y_\theta \sim 8$ and $P(y_\theta \ge 8) \sim 0.2$. On the other hand, if $\theta_{12,0} = 1$ deg., then $y_\theta \sim 27$, corresponding to $P(y_\theta \ge 27) \sim 0.12$. Recall that we are using empirical values for $P(y)$ derived from simulations, since Eq. (\ref{eq:f_y_CDF}) breaks down when $m_1 \sim m_2$. To conclude, the observed mutual inclination in the system can be easily generated by ``1+2'' scattering if $\theta_{12,0} \gtrsim 3$ deg., and is still possible with $P \sim 0.12$ probability for $\theta_{12,0} \sim 1$ degree.
\end{itemize}

In summary, we have found that each of the systems HAT-P-11, Gliese 777 A and $\pi$ Men have inner planet eccentricities and mutual inclinations that are inconsistent with being produced by secular forcing from their external perturber alone, but is consistent with the ``1+2'' scattering hypothesis ($p > 0.10$ in all cases). In addition, direct scatterings of the inner planet by the outer giants during ``1+2'' scattering could under certain regimes produce additional excitation in eccentricity and inclination, which further bolsters the prospects the currently observed eccentricities and mutual inclinations being explained by ``1+2'' scattering.

\section*{Acknowledgements}
We thank the anonymous referee for helpful comments that improved the manuscript. BP is supported by the NASA Earth and Space Sciences Fellowship. DL thanks the Dept. of Astronomy and the Miller Institute for Basic Science at UC Berkeley for hospitality while part of this work was carried out.

\section*{Data Availability}
Original data produced by our simulations in this work are available upon request.

\bibliography{msNotes}
\bibliographystyle{mnras}

\appendix
\section{Calculation of Moments of $\mathcal{E}_a$ }
\label{sec:A1}
We demonstrate how to calculate the various moments of an inner planet subject to a stochastic secular forcing.
For case 1, the unconstrained perturber, from Eq. (\ref{eq:ea}) the mean of $\mathcal{E}_a$ is given by
\begin{align}
    \langle \mathcal{E}_a \rangle &= \Big \langle \int_0^{t_{\mathrm{ej}}} e^{i \omega_{a1} (s-t_{\mathrm{ej}})} i \nu_{a1} Z(s) ds  \Big \rangle \nonumber \\ 
    &= \int_0^{t_{\mathrm{ej}}} e^{i \omega_{a1} (s-t_{\mathrm{ej}})} i \nu_{a1}  \langle Z(s)   \rangle ds  = 0. 
\end{align}
The variance of $\mathcal{E}_a$ is 
\begin{align}
   \langle | \mathcal{E}_a |^2 \rangle &= \Big \langle \left| \int_0^{t_{\mathrm{ej}}} e^{-i \omega_{a} (s-t_{\mathrm{ej}})} i \nu_{a1} Z(s) ds \right| ^2  \Big \rangle \nonumber \\
   &= \nu^2_{a1} \Big \langle \left(\int_0^{t_{\mathrm{ej}}} e^{ii \omega_{a1} (s-t_{\mathrm{ej}})} Z(s) ds  \right) {\left(\int_0^{t_{\mathrm{ej}}} e^{i \omega_{a1} (r-t_{\mathrm{ej}})} Z^*(r) dr  \right)} \Big \rangle \nonumber \\
   &= \nu^2_{a1} \left(\int_0^{t_{\mathrm{ej}}} \int_0^{t_{\mathrm{ej}}} e^{i \omega_{a1} (r-s)} \langle Z(s) Z^*(r) \rangle  ~ds ~dr \right)  \nonumber \\ 
   &= 2\sigma^2_{\mathcal{E}1}  \nu^2_{a1} \left(\int_0^{t_{\mathrm{ej}}} \int_0^r e^{i \omega_{a} (r-s)} s  ds ~dr 
   + \int_0^{t_{\mathrm{ej}}} \int_r^{t_{\mathrm{ej}}} e^{i \omega_{a1} (r-s)} r  ds ~dr  \right)  \nonumber  \\
   &= 4 \left(\frac{\nu_{a1}}{\omega_{a}} \right)^2 \left[1 - \frac{\sin{(\omega_{a}t_{\mathrm{ej}})}}{\omega_{a}t_{\mathrm{ej}}}\right] \sigma^2_{\mathcal{E}1} t_{\mathrm{ej}}.  
\end{align}
Similarly the covariance between $\mathcal{E}_{a,\mathrm{ej}}$ and its forced eccentricity is given by
\begin{align}
    \langle \mathrm{Re}(\mathcal{E}_{a,\mathrm{ej}}\mathcal{E}_{a,\mathrm{forced}}^*) \rangle &= \Big \langle \mathrm{Re} \left(\int_0^{t_\mathrm{ej}} -i \nu_{a1} e^{-i \omega_{a1}(s-t_{\mathrm{ej}})} Z(s) \frac{\nu_{a1}}{\omega_{a1}} Z^*(s) ds \right)     \Big \rangle \nonumber \\ 
    &= \mathrm{Im} \left(\int_0^{t_\mathrm{ej}} \nu_{a1} e^{-i \omega_{a1}(s-t_{\mathrm{ej}})} \frac{\nu_{a1}}{\omega_{a1}}  \langle Z(s) Z^*(s) \rangle ~ds \right)  \nonumber \\
    &= 2~ \mathrm{Im} \left(\int_0^{t_\mathrm{ej}} \nu_{a1} e^{-i \omega_{a}(s-t_{\mathrm{ej}})} \frac{\nu_{a1}}{\omega_{a1}} \sigma^2_{\mathcal{E}1} s ~ds \right)     \nonumber \\
    &= 2 \left(\frac{\nu_{a1}}{\omega_{a1}} \right)^2 \left[1 - \frac{\sin{(\omega_{a1}t_{\mathrm{ej}})}}{\omega_{a1}t_{\mathrm{ej}}}\right] \sigma^2_{\mathcal{E}1} t_{\mathrm{ej}}.
\end{align}
The case of the constrained perturber (Brownian bridge) is analogous to the case for the unconstrained perturber, except with $Z(s) \rightarrow B(s)$. The expectations of $B(s)$ are given by Eqs. (\ref{eq:mean_B}) - (\ref{eq:covar_B}). 

\bsp	
\label{lastpage}

\end{document}